\documentclass[preprint,review,12pt]{elsarticle}

\usepackage{hyperref}
\usepackage{amsmath}
\usepackage{bm}
\usepackage{amssymb}
\usepackage{amsthm}
\usepackage{array}
\usepackage{enumitem}
\usepackage{graphicx}
\usepackage{subcaption}

\usepackage{pgfplots}
\pgfplotsset{compat=newest}
\usepgfplotslibrary{external}
\usepgfplotslibrary{patchplots}
\usetikzlibrary{shapes,arrows}
\usetikzlibrary{positioning,calc}
\usetikzlibrary{decorations.pathreplacing,decorations.markings,math,arrows.meta}
\usetikzlibrary{spy}
\usetikzlibrary{pgfplots.groupplots}

\usepackage[ruled,vlined,linesnumbered]{algorithm2e}

\usepackage{geometry}
\journal{Elsevier}

\newcommand\figureFontSize{8}

\begin{document}

\begin{frontmatter}

\title{An adapted deflated conjugate gradient solver for robust extended/generalised finite element solutions of large scale, 3D crack propagation problems.}

\author[1]{Konstantinos Agathos}
\author[1,2]{Tim Dodwell}
\author[3]{Eleni Chatzi}
\author[4,5]{St\'{e}phane P. A. Bordas}

\address[1]{College of Engineering Mathematics and Physical Sciences, University of Exeter, UK}
\address[2]{The Alan Turing Institute, UK}
\address[3]{Department of Civil, Environmental, and Geomatic Engineering, ETH Z\"{u}rich, Switzerland}
\address[4]{Research Unit in Engineering Science, Luxembourg University, Luxembourg}
\address[5]{Institute of Theoretical, Applied and Computational Mechanics, Cardiff University, UK}




\begin{abstract}
An adapted deflation preconditioner is employed to accelerate the solution of linear systems resulting from the discretization of fracture mechanics problems with well-conditioned extended/generalized finite elements. The deflation space typically used for linear elasticity problems is enriched with additional vectors, accounting for the enrichment functions used, thus effectively removing low frequency components of the error. To further improve performance, deflation is combined, in a multiplicative way, with a block-Jacobi preconditioner, which removes high frequency components of the error as well as linear dependencies introduced by enrichment. The resulting scheme is tested on a series of non-planar crack propagation problems and compared to alternative linear solvers in terms of performance.
\smallskip
\end{abstract}


\end{frontmatter}


\section{Introduction}

The eXtended and Generalised Finite Element Methods (XFEM and GFEM respectively) \cite{Moes1999,Strouboulis2000}, form a class of multiscale finite element methods that have significantly contributed to the increase of the level of automation and computational efficiency of crack propagation simulations. These approaches substantially reduce, or even eliminate, the need for re-meshing, allowing small-scale features to be captured efficiently and accurately at a coarse, marco-scale representation. This is accomplished by means of Partition of Unity (PoU) enrichment~\cite{Melenk1996}, whereby polynomial approximation spaces of the Finite Element Method (FEM) are supplemented with appropriate additional enrichment functions. Importantly, these enrichment functions encompass known features or micro-scale information of the system. In fracture mechanics, features such as discontinuities and singularities can be represented independently of the underlying Finite Element (FE) mesh, offering significant computational advantages. This has contributed to the popularity of the methods, and has led to their use in a vast array of applications, now documented in numerous review papers~\cite{Belytschko2009,Fries2010,Sukumar2015,Egger2019}. 

\smallskip
The advantages of the method do not come without a price. Among other challenges, the method can produce ill-conditioned systems of equations, as a result of linear dependencies introduced by enrichment. In the case of large scale, three dimensional simulations, for which parallel iterative solvers are a necessity, this ill-conditioning, without special attention, leads to stagnation (often complete failure) of the solver. To mitigate these issues, several approaches have been proposed, seeking to resolve the linear dependencies introduced by the enrichment functions. For the discontinuous functions used to represent solutions along crack faces \cite{Moes1999}, strategies include the elimination of the pathological enriched degrees of freedom \cite{Lang2014}, and particular stabilisation techniques either at the element \cite{Loehnert2014} or system level \cite{Ventura2016}. For enrichment with asymptotic functions, typically used to represent singularities occurring at crack tips/fronts, alternative enrichment functions have been proposed \cite{Duarte2000,Chevaugeon2013}, alongside enrichment modification strategies, such as degree of freedom gathering \cite{Laborde2005,Agathos2015,Agathos2016}, the so-called stable GFEM \cite{Babuvska2012,Gupta2015,Sanchez2019} and orthogonalisation \cite{Agathos2019-a,Agathos2019-b}.


\smallskip
Most of the above strategies seek to improve the conditioning of the resultant system of equations by acting directly on the formulation of the XFEM/GFEM methodology and the choice, selection or representation of the enrichment functions. Stepping back from this specific XFEM/GFEM setting, to the more general challenge of parallel iterative solvers for ill-conditioned systems of equations, a significant amount of work exists on robust preconditioners for parallel iterative solvers. For a general ill-conditioned system ${\bf K}{\bf u} = {\bf f}$, the idea of (left) preconditioning is to determine an operator ${\bf M}^{-1}$ for which ${\bf M}^{-1} {\bf K}$ is well-conditioned and ${\bf M}^{-1}$ is efficient to construct; allowing a parallel iterative solver, e.g. Conjujate Gradient (pCG), to converge quickly. The most widely used preconditioners for iterative solvers in both commercial and scientific FE codes are Algebraic Multigrid (AMG) methods \cite{Henson2002}. They have demonstrated excellent scalability for a broad class of problems over thousands of processors, and have the advantage of operating exclusively on the matrix equations, so that they can be applied as a ‘black-box’. As a preconditioner, AMG constructs the matrix ${\bf M}$ by repeatedly coarsening the full matrix ${\bf K}$ through recursive aggregation over the degrees of freedom. The aggregation process is algebraic and based on the fact that the solution at two neighbouring nodes will be similar if they are ‘strongly connected’. The success of an AMG preconditioner depends on this aggregation process. In the vicinity of a crack, and with large simulation divided into connected subdomains, this aggregration can be difficult to design properly. A widely used alternative is additive Schwarz based preconditioners. Single level versions utilise independent solves of the local Dirichlet problem. These are computed in parallel, using robust direct solvers, with the aim of resolving the ill-conditioning of the problem at a local length scale. These can be particularly effective, yet are known to lack robustness in the case of large numbers of subdomains or with respect to large contrast in coefficients. A problem ever present in crack propagation problems. The problems in this case are linked to the fact that the localised solves do not resolve the global low energy modes of the system, meaning the ill-conditioning persists. To overcome this issue, preconditioners can be augmented with an additional step, which involves either deflation methods \cite{Toselli2004}, or second level coarse-solvers. In the former case, the deflection vectors/ coarse basis for the problem can be identified through an analytical solution (e.g. ZEM \cite{Toselli2004}), while in the latter, optimal approximations the global low energy modes are identified via local eigenproblems, see for example methods involving Generalised Eigenproblems in the Overlap GenEO \cite{Spillane2014,Reinarz2018,Butler2020}. It is interesting to note that construction of robust coarse spaces, via GenEO type methods bears strong ties to Generalised Finite Element Methods, as studied by Ma et al.~\cite{Ma2021}. In some respect, this optimal basis is an optimal presentation of enrichment functions, which dooes not suffer from linear dependencies. Yet with this robustness comes the additional computational burden associated to solution of the local eigenvalue problems, which in many cases is an overkill.

\smallskip
Several techniques have also been proposed to improve the performance of iterative solvers for crack propagation problems discretised using the XFEM/GFEM. The preconditioners proposed by B\'{e}chet et al.~\cite{Bechet2005} and Menk and Bordas~\cite{Menk2011} employ a Cholesky decomposition to remove linear dependencies introduced by enrichment. The application of AMG preconditioners to systems produced by the XFEM/GFEM has also been proposed in several works. However, as shown, for instance, in Berger-Vergiat et al.~\cite{Berger2012}, it cannot be performed directly since it results in sub-optimal performance. This is attributed to the fact that discontinuities present along crack faces are not accounted for in the construction of prolongation/restriction operators. To overcome this limitation, some works \cite{Berger2012,Waisman2013,Svolos2020} employ Schwartz preconditioners to decompose the problem domain into a healthy and a cracked part, for which different solvers, such as AMG or LU, can be used. It should be noted that this type of decomposition has also been exploited for other purposes in the XFEM/GFEM literature, for instance in the seminal work of Bordas and Moran \cite{Bordas2006,Bordas2007} it is used to enable the combination of commercial and research codes, while in the global-local GFEM \cite{Duarte2008} it is employed to construct numerically derived enrichment functions. Other works \cite{Hiriyur2012,Gerstenberger2013} modify prolongation/restriction operators to account for discontinuities along crack surfaces. In a more recent work~\cite{Feng2019}, the block structure of linear systems produced by the XFEM/GFEM is exploited to derive a modified system, where AMG and reanalysis techniques can be applied to accelerate the solution. This block structure is also exploited in the work of Fillmore et al.~\cite{Fillmore2017} through specialised block-Jacobi and block-Gauss-Seidel preconditioners. 

\smallskip
In this work, a preconditioner based on adapted deflation~\cite{Smith2004} is proposed for crack propagation problems, discretised with the XFEM/GFEM. The deflation space is constructed based on domain partitioning \cite{Aubry2011}, while additional deflation vectors are introduced to account for the discontinuities introduced by the cracks. The approach is combined with a block-Jacobi preconditioner, which offers several advantages such as high efficiency, straightforward parallelisation, and efficient update for crack propagation problems. Whilst the main target application is non-planar crack propagation, the approach would also be suitable for other types of problems, for instance inverse problems~\cite{Agathos2018-b}, where repeated solutions of linear systems with slight variations, introduced by different crack configurations, are required.

\smallskip
The remainder of the paper is structured as follows: after a brief presentation of the elasticity equations and adopted discretisation scheme in \autoref{problem statement}, the different components of the proposed preconditioner are described in \autoref{adapted delfation preconditioner}, and tested in \autoref{numerical examples} through a series of examples involving crack propagation in models of components of increasing geometrical complexity and size. Finally, in \autoref{conclusion}, the results are summarised and conclusions are drawn.
\section{Problem statement and discretization} \label{problem statement}

\subsection{Weak form}

\begin{figure}
	\centering
		\includegraphics[width=0.5\textwidth]{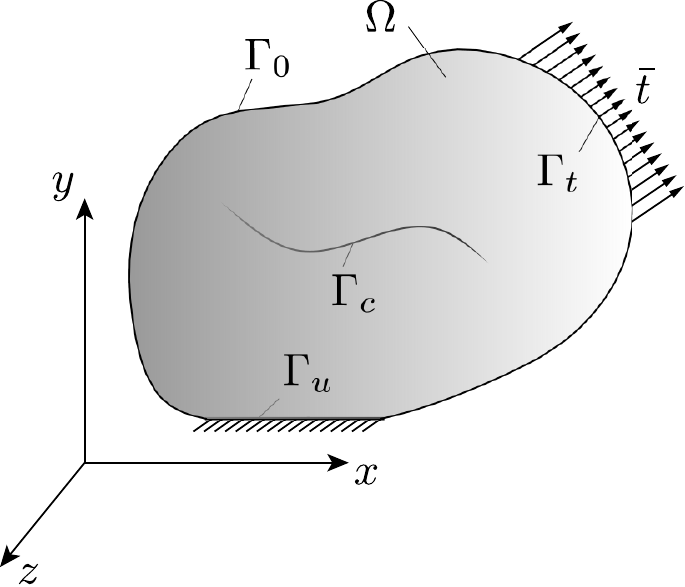}
	\caption{Cracked body and boundary conditions.}
	\label{fig:3dbody}
\end{figure}

For linear elastic fracture mechanics, we consider the linear elasticity problem for a cracked domain $\Omega$, whose boundary can be decomposed, as illustrated in \autoref{fig:3dbody}, in a part where free surface conditions apply ($\Gamma_0$), a part where displacements $\bar{\bm{u}}$ are imposed as Dirichlet conditions ($\Gamma_u$), a part where surface tractions $\bar{\bm{t}}$ are applied ($\Gamma_t$) and the crack faces $\Gamma_c$, where free surface conditions also apply. The weak form of the problem, assuming a linear elastic constitutive equation, can be written as:

\noindent Find $\bm{u} \in \mathcal{U}$ such that $\forall \bm{v} \in \mathcal{V}^0$:

\begin{equation}\label{eq:weak_form_const_eq}
	\int_{\Omega} \bm{\epsilon} (\bm{u}) : \bm{D} : \bm{\epsilon} (\bm{v}) \ d\Omega = \int_{\Omega} \bm{b}\cdot\bm{v} \ d\Omega + \int_{\Gamma_t} \bar{\bm{t}}\cdot\bm{v} \ d\Gamma
\end{equation}

\noindent with :

\begin{equation}
	\mathcal{U} = \left\{\bm{u}|\bm{u} \in \left(H^1\left(\Omega\right)\right)^3,\bm{u} = \bar{\bm{u}}  \text{ on }  \Gamma_u \right\}
\end{equation}

\noindent and

\begin{equation}
	\mathcal{V}^0 = \left\{\bm{v}|\bm{v} \in \left(H^1\left(\Omega\right)\right)^3,\bm{v} = \bm{0}  \text{ on }  \Gamma_u \right\}
\end{equation}

\noindent In \autoref{eq:weak_form_const_eq}, $\bm{u}$ is the displacement field, $\bm{v}$ is the virtual displacement field, $\bm{D}$ is the Hooke tensor, $\bm{\sigma}$ is the Cauchy stress tensor and $\bm{\epsilon}$ is the linear strain tensor. 

\subsection{Crack representation}

As typically done within the XFEM/GFEM framework, cracks are implicitly represented using the level set method. To this end, two signed distance functions are defined:

\begin{itemize}[topsep=0.0pt]
  \item The normal level set $\phi$, defined as the signed distance from the crack surface:
	\begin{equation}\label{eq:sgned_distance}
		\phi\left(\bm{x}\right)= \min_{\bar{\bm{x}}\in\Gamma_c}{\left\|\bm{x}-\bar{\bm{x}}\right\|\text{sign}\left(\bm{n}^+\cdot\left(\bm{x}-\bar{\bm{x}}\right)\right)} 
	\end{equation}
	where $\bm{n}^+$ is the outward normal to the crack surface and $\text{sign}\left( \square \right)$ is the $\text{sign}$ function.
	\item The tangential level set $\psi$, defined as the signed distance function satisfying the conditions:
	    \begin{subequations}
	        \begin{equation}
	            \nabla\phi\cdot\nabla\psi=0
	        \end{equation}
	        \begin{equation}
	        \left.
	            \begin{array}{c}
	                \phi\left(\bm{x}\right)=0 \\
	                \psi\left(\bm{x}\right)=0
	            \end{array}
	            \right\} \forall \bm{x} \in \Gamma_f
	        \end{equation}
	    \end{subequations}
    where $\Gamma_f$ are the crack front/tips.
\end{itemize}

Based on these functions, a polar coordinate system, with its origin at the crack front, can be defined:

\begin{equation}\label{eq:polar_coords}
	r=\sqrt{\phi^2+\psi^2}, \qquad \theta=\arctan\left(\frac{\phi}{\psi}\right)
\end{equation}

In general, the integration of several evolution equations is required to update level set descriptions of propagating cracks \cite{Duflot2007}, a procedure that can be both complex and computationally demanding. Herein, the vector level set method~\cite{Ventura2003,Agathos2017,Agathos2018-a} is employed, which allows to update the crack description using only geometrical operations, thus simplifying the process.

\subsection{Discretization}

The weak form of \autoref{eq:weak_form_const_eq} is discretized using the standard XFEM displacement approximation with shifted jump enrichment functions and quasi-orthogonalized tip enrichment functions:

\begin{equation}\label{eq:xfem_displ}
	\bm{u}^h\left(\bm{x}\right) =\sum_{\forall I \in \mathcal{N}} N_I\left(\bm{x}\right) \bm{u}_I +\sum_{\forall J \in \mathcal{N}^j} N_J\left(\bm{x}\right) H_J\left(\bm{x}\right) \bm{b}_J
+\sum_{\forall T\in \mathcal{N}^t} \sum_{\forall j} N_T\left(\bm{x}\right) \bar{F}_{Tj}\left(\bm{x}\right) \bm{c}_{Tj}
\end{equation}

\noindent where $N_I$ are the FE interpolation functions, $\bm{u}_I$ are FE degrees of freedom (dofs), and $\bm{b}_J$, $\bm{c}_{TJ}$ are the enriched degrees of freedom. The shifted jump enrichment functions are defined as:

\begin{equation}
    H_J\left(\bm{x}\right) = H\left(\bm{x}\right) - H\left(\bm{x}_J\right)
\end{equation}

\noindent where $\bm{x}_J$ are the nodal coordinates of node $J$ and:

\begin{equation}\label{eq:jump_enr_functions}
	H\left(\bm{x}\right) = H(\phi\left(\bm{x}\right))=\left\{
	\begin{array}{lr}
     &1  \quad  \text{for} \ \phi \geq 0 \\
    -&1  \quad  \text{for} \ \phi < 0
  \end{array}\right.
\end{equation}

The quasi-orthogonalized tip enrichment functions are defined as in one of the previous works of the authoring team~\cite{Agathos2019-a}:

\begin{align}
	&\bar{\bm{F}}_{I}\left(r,\theta\right)& =  \left\{ \right. & \left. \sqrt{r}  \sin \frac{\theta}{2}- \sqrt{r_I} \sin \frac{\theta_I}{2}, \right. \nonumber \\
   & & & \left. \sqrt{r} \cos \frac{\theta}{2}- \sqrt{r_I} \cos \frac{\theta_I}{2}, \right. \nonumber \\ & &  & \left. 2 \sqrt{r} \cos \left( \frac{\theta}{2} + \theta_I \right) \sin \left( \frac{\theta-\theta_I}{2} \right), \right. \nonumber \\ & & & \left. -2 \sqrt{r} \sin^2 \left( \dfrac{\theta-\theta_I}{2} \right) \left[ c_{I34} \cos \left( \frac{\theta}{2} + \theta_I \right) + \sin \left( \frac{\theta}{2} + \theta_I \right) \right]  \right\}
   \label{eq:orthogonal_tip_enr_func}
\end{align}

\noindent where $r_I, \theta_I$ are the nodal values of the radius $r$ and angle $\theta$ and:

\begin{equation}
    c_{I34}=-\dfrac{7 \sin \left(\theta_I\right)+11 \sin \left( 2 \theta_I \right) + 65 \sin \left( 3 \theta_I \right)}{36 \left[1+\cos \left(3 \theta_I \right) \right]}
\end{equation}

The nodal sets in \autoref{eq:xfem_displ} are defined as:

\begin{description}[style=multiline,leftmargin=0.75cm]
\item[$\mathcal{N}$] is the set of all nodes in the FE mesh.
\item[$\mathcal{N}^j$] is the set of jump enriched nodes. This nodal set includes all nodes whose support is split in two by the crack.
\item[$\mathcal{N}^t$] is the set of tip enriched nodes. This nodal set includes all nodes whose support includes the crack front or nodes with $r_I \leq r_e$, with $r_e$ a user defined enrichment radius.
\end{description}

It should be mentioned that the above choice of enrichment functions, combined with linear tetrahedral elements, was shown to lead to well conditioned system matrices~\cite{Agathos2019-a}.

\subsection{Linear system of equations}\label{linear system of equations}

Substituting the displacement approximation of \autoref{eq:xfem_displ} into the weak form of \autoref{eq:weak_form_const_eq} and carrying out the integrations, yields a system of equations with the following block structure:

\begin{equation}
    \underbrace{
    \left[	
	\begin{array}{c c }
    	\bm{K_{FF}}   & \bm{K_{FX}}  \\
     	\bm{K_{FX}}^T & \bm{K_{XX}}
  \end{array}\right]}_{\bm{K}} \underbrace{\left[	
	\begin{array}{c}
    	\bm{u_{F}} \\
     	\bm{u_{X}}
  \end{array}\right]}_{\bm{u}} = \underbrace{\left[	
	\begin{array}{c}
    	\bm{f_{F}} \\
     	\bm{f_{X}}
  \end{array}\right]}_{\bm{f}}
    \label{eq:linear_system}
\end{equation}

\noindent where $\bm{K} \in \mathbb{R}^{n \times n}$ is the stiffness matrix, $\bm{u} \in \mathbb{R}^n$ is the vector of unknowns, $\bm{f} \in \mathbb{R}^n$ is the load vector, and $n$ is the total number of unknowns. Moreover, $\bm{u_F} \in \mathbb{R}^{n_F}$ contains all the standard degrees of freedom, corresponding to $\bm{u}_{I}$ from \autoref{eq:xfem_displ}, $\bm{u_X} \in \mathbb{R}^{n_X}$ contains all the enriched degrees of freedom, corresponding to $\bm{b}_J$ and $\bm{c}_{Tj}$ from \autoref{eq:xfem_displ}, $\bm{K_{FF}} \in \mathbb{R}^{n_F \times n_F}$, $\bm{K_{FX}} \in \mathbb{R}^{n_F \times n_X}$, $\bm{K_{X}} \in \mathbb{R}^{n_X \times n_X}$ are the parts of the stiffness matrix corresponding to the standard and enriched degrees of freedom, as well as their interaction, $\bm{f_{F}} \in \mathbb{R}^{n_F}$, $\bm{f_{X}} \in \mathbb{R}^{n_X}$ are the parts of the load vector corresponding to the standard and enriched degrees of freedom, $n_F$ is the number of standard degrees of freedom, $n_X$ is the number of enriched degrees of freedom. From the above it should be obvious that $n = n_F + n_X$.

While \autoref{eq:linear_system} does not necessarily reflect the order in which equations are stored in practice, it helps gain some insight regarding the structure of the resulting systems. For instance, $\bm{K_{FF}}$ will typically contain more than 90\% of the total entries in the stiffness matrix and will remain constant during a crack propagation simulation, with only the other three blocks changing as cracks propagate and additional nodes are enriched.
\section{Adapted deflation preconditioner} \label{adapted delfation preconditioner}

\subsection{The conjugate gradient method} \label{the conjugate gradient method}

For two dimensional or small three dimensional problems, the linear system of \eqref{eq:linear_system} can be solved efficiently using direct solvers. However, for larger three dimensional problems, where direct solvers can be inefficient due to excessive memory requirements, iterative solvers become an attractive alternative. For symmetric systems, the most widely used alternative is the conjugate gradient method. As shown in \autoref{alg:CG}, the method iteratively improves upon an initial solution $\bm{u}_{0}$ by adding vectors $\bm{p}_{i}$, which are conjugate with respect to $\bm{K}$. Since the number of iterations required to obtain an accurate solution depends on the conditioning of $\bm{K}$, a preconditioner $\bm{M}$ is typically employed, the choice of which can significantly affect the efficiency of the solver. In what follows, a deflation based preconditioner, specifically tailored for systems of the form of \eqref{eq:linear_system} is presented.

\begin{algorithm}
\SetAlgoLined
\KwData{$\bm{K}$, $\bm{f}$, $\bm{u}_{0}$, $\bm{M}$}
\KwResult{$\bm{u}_{i}$ }

    $\bm{r}_{0}  = \bm{f} - \bm{K} \bm{u}_{0}$\;
    $\bm{z}_{0}  = \bm{M}^{-1} \bm{r}_{0}$\;
    $\bm{p}_{0} = \bm{z}_{0}$
    
    \For{$i=0,1,2,\dots$}{
        $\alpha_{i} = \dfrac{ \bm{r}_i^T \bm{z}_i}{ \bm{p}_i^T \bm{K} \bm{p}_i}$;
        
        $\bm{u}_{i+1} = \bm{u}_{i} + \alpha_{i} \bm{p}_{i}$
        
        $\bm{r}_{i+1} = \bm{r}_{i} - \alpha_{i} \bm{K} \bm{p}_{i}$
        
        $\bm{z}_{i+1} = \bm{M}^{-1} \bm{r}_{i+1}$
        
        $\beta_{i} = \dfrac{\bm{r}_{i+1}^T \bm{z}_{i+1}}{ \bm{r}_{i}^T \bm{z}_{i}}$
        
        $\bm{p}_{i+1} = \bm{z}_{i+1} + \beta_{i} \bm{p}_{i}$
    }
    
    \caption{The conjugate gradient method.}
    \label{alg:CG}
\end{algorithm}

\subsection{Deflation} \label{deflation}

Given a basis for a subspace of $\mathbb{R}^{n \times n}$, deflation \cite{Nicolaides1987,Saad2000} aims at accelerating the iterative solution of linear systems, such as the one of \autoref{eq:linear_system}, by solving a modified system, of the form:

\begin{equation}
    \hat{\bm{K}} \hat{\bm{u}} = \hat{\bm{f}}
    \label{eq:deflated_system}
\end{equation}

\noindent where:

\begin{subequations}
    \begin{equation}
        \hat{\bm{K}} = \bm{P}^T\bm{K}
    \end{equation}
    \begin{equation}
        \hat{\bm{f}} = \bm{P}^T\bm{f}
    \end{equation}
    \begin{equation}
        \bm{P} = \bm{I} - \bm{K} \bm{Q}
    \end{equation}
    \begin{equation}
    \bm{Q} = \bm{W} \left( \bm{W}^T\bm{K}\bm{W}\right)^{-1}\bm{W}^T
\end{equation}
\end{subequations}

\noindent with $\bm{W} \in \mathbb{R}^{n \times k}$ a matrix containing a basis of a subspace of $\mathbb{R}^{n}$, with $k \ll n$. It can be shown that if $\bm{W}$ is full rank, and $\bm{K}$ is symmetric positive definite (SPD), the system of \autoref{eq:deflated_system} can be solved using the conjugate gradient (CG) method, while the solution for the original system can be recovered as:

\begin{equation}
    \bm{u} = \bm{Q} \bm{f} + \bm{P}^T \hat{\bm{u}}
\end{equation}

In \autoref{eq:deflated_system}, $\bm{P}$ projects the subspace defined by $\bm{W}$ out of the original system. While it can be shown that any full rank matrix will result in some reduction of the number of iterations required by the CG method, typically, columns of $\bm{W}$ are selected to approximate eigenvectors corresponding to the smallest eigenvalues of $\bm{K}$. Since the projection of \autoref{eq:deflated_system} essentially removes the corresponding eigenvalues of $\bm{K}$, such a choice effectively reduces its condition number, resulting in a reduced number of CG iterations. More details about the construction of $\bm{W}$ for the problems considered herein are given in \autoref{deflation space}.

Deflation is also related to multi-grid methods, more specifically, $\bm{W}^T$ and $\bm{W}$ can be seen as prolongation and restriction operators respectively \cite{Tang2007}. Then, $\left( \bm{W}^T\bm{K}\bm{W}\right)$ can be viewed as a coarse grid matrix and the application of $\bm{Q}$ to the residual during the iterative solution of the system represents a coarse grid correction, aiming at removing low frequency components of the error.

Deflation, viewed either as a projection or a coarse grid correction, can be combined to preconditioning in an additive or multiplicative way, resulting in multiple alternatives. For instance, applying deflation to a system preconditioned with an arbitrary SPD preconditioner $\bm{M}^{-1}$ results in the deflated CG algorithm introduced by Nicolaides~\cite{Nicolaides1987}.

In Tang et al.~\cite{Tang2007}, a comparison of several alternatives was conducted, and it was concluded that a multiplicative combination of a general preconditioner with the coarse grid correction $\bm{Q}$, as discussed, for instance, in Smith et al.~\cite{Smith2004}, results in the most robust and efficient method. The combination, termed ``Adapted Deflation Variant 2'' (A-DEF2) results in a preconditioner of the form:

\begin{equation}
    \bm{M}_{\text{A-DEF2}} = \bm{P}^T\bm{M}^{-1}+\bm{Q}
\end{equation}

\noindent where $\bm{M}^{-1}$ can be an arbitrary preconditioner. The operator is not SPD, however it was shown to be equivalent to other symmetric operators, including the one corresponding to the deflated CG~\cite{Nicolaides1987}, provided that the following initial solution is used:

\begin{equation}
    \bm{u_0} = \bm{Q} \bm{f} + \bm{P}^T \bm{u}_{-1}
\end{equation}

\noindent where $\bm{u}_{-1}$ is some arbitrary initial solution vector.

In \autoref{alg:A-DEF2_u0}, the efficient computation of this initial solution is described, while in \autoref{alg:A-DEF2} the steps required to efficiently apply A-DEF2 as a preconditioner, given a residual $\bm{r}$ are described. It should be mentioned that, both of these can be used in conjunction to the standard CG method, described in \autoref{alg:CG}.

\begin{algorithm}
\SetAlgoLined
\KwData{$\bm{K}$, $\bm{W}$, $\bm{f}$, $\bm{u}_{-1}$}
\KwResult{$\bm{u}_{0}$ }

    Solve: $\bm{W}^T \bm{K} \bm{W} \bm{\lambda}  = \bm{W}^T \bm{K} \bm{u}_{-1}$\;
    Solve: $\bm{W}^T \bm{K} \bm{W} \bm{\mu}  = \bm{W}^T \bm{f}$\;
    
    $\bm{u}_{0} = \bm{u}_{-1} + \bm{W} \left( \bm{\mu} - \bm{\lambda} \right)$\;
    \caption{Initial solution to be used with the A-DEF2 preconditioner.}
    \label{alg:A-DEF2_u0}
\end{algorithm}

\begin{algorithm}
\SetAlgoLined
\KwData{$\bm{K}$, $\bm{M}$, $\bm{W}$, $\bm{r}$}
\KwResult{$\bm{z}$ }
    $\bm{y} = \bm{M}^{-1} \bm{r}$;

    Solve: $\bm{W}^T \bm{K} \bm{W} \bm{\mu}  = \bm{W}^T \bm{K} \bm{y}$\;

    Solve: $\bm{W}^T \bm{K} \bm{W} \bm{\lambda}  = \bm{W}^T \bm{r}$\;
    $\bm{z} = \bm{y} + \bm{W} \left(  \bm{\lambda} - \bm{\mu} \right)$\;
 
    \caption{A-DEF2 preconditioner.}
    \label{alg:A-DEF2}
\end{algorithm}

\subsection{Deflation space}\label{deflation space}

The choice of deflation space is critical to the performance of deflation-based preconditioners. To this end, different approaches have been proposed to efficiently approximate lower eigenvalues of linear systems resulting from the discretisation of different problems. Extending the categorisation of Diaz-Cortes et al.~\cite{Cortes2018}, the following techniques can be identified for constructing deflation spaces:

\begin{description}
    \item[Recycling deflation] In this technique, the most effective vectors of Krylov-subspaces used in previous solutions are identified and re-used as deflation vectors \cite{Parks2006}.
    \item[Subdomain deflation] Using this approach, the problem domain is divided into subdomains and vectors corresponding to a constant solution in each subdomain are used to form the deflation space \cite{Nicolaides1987}.
    \item[Multigrid and multilevel deflation] As mentioned in \autoref{deflation}, deflation spaces can also be seen as prolongation/restriction operators \cite{Tang2007}. Thus prolongation/restriction operators constructed for multigrid/multilevel methods can also be used as deflation vectors.
    \item[POD based deflation] Using this approach, solutions from problems similar to the problem to be solved are compressed using the Proper Orthogonal Decomposition (POD) and used as deflation vectors \cite{Cortes2018}. It should be mentioned that the POD can also be used in combination to Krylov solvers in reduced order modelling methods, such as the ones proposed by Kerfriden et al. \cite{Kerfriden2011,Kerfriden2013}, where a limited number of CG iterations are employed to enrich lower dimensional spaces used for fracture problems.
\end{description}

Herein we employ subdomain deflation, while we enrich the deflation space with additional vectors to account for cracks.

\subsubsection{Deflation vectors for linear elasticity}\label{deflation space elasticity}

For general linear elasticity problems, subdomain-based deflation spaces can be constructed by considering rigid body translations and rotations of each subdomain~\cite{Aubry2011}. This approach results in deflation matrices assuming the following values for each node of the subdomain of interest:

\begin{equation}
	\bm{W}_I =\left[	
	\begin{array}{c c c c c c}
    	1 & 0 & 0 &  0   &  z_I & -y_I  \\
     	0 & 1 & 0 & -z_I &  0   &  x_I  \\
        0 & 0 & 1 &  y_I & -x_I &  0
  \end{array}\right]
  \label{eq:dcg_deflation_space}
\end{equation}

\noindent where $x_I, y_I, z_I$ are the spatial coordinates of node $I$.

To render deflation effective, the problem domain should be divided evenly and subdomains should also be contiguous. In our implementation, METIS~\cite{Karypis1997} is used for this task, which offers additional features, that, as described in \autoref{preconditioning}, can be used to further accelerate the solution.


\subsubsection{Enriched deflation vectors for XFEM/GFEM}

For enriched approximations, we introduce a set of additional deflation vectors to account for the presence of the crack. These additional modes are only defined in subdomains containing enriched nodes and are derived by considering modes of deformation corresponding to the enrichment functions. For instance, for the modified Heaviside enrichment function the following modes are considered:

\begin{align}
    \label{eq:enriched_deflated_modes}
    \bm{u}^h\left(\bm{x}_I\right) = \left[ \begin{array}{c}
    	H\left(\bm{x}_I\right) \\
     	0 \\
        0 \\
  \end{array}\right], 
  \bm{u}^h\left(\bm{x}_I\right) = \left[ \begin{array}{c}
    	0 \\
     	H\left(\bm{x}_I\right) \\
        0 \\
  \end{array}\right], 
  \bm{u}^h\left(\bm{x}_I\right) = \left[ \begin{array}{c}
    	0 \\
     	0 \\
        H\left(\bm{x}_I\right) \\
  \end{array}\right] \\
  \bm{u}^h\left(\bm{x}_I\right) = \left[ \begin{array}{c}
    	0 \\
     	- H\left(\bm{x}_I\right) z_I \\
        H\left(\bm{x}_I\right) y_I \\
  \end{array}\right], 
  \bm{u}^h\left(\bm{x}_I\right) = \left[ \begin{array}{c}
    	H\left(\bm{x}_I\right) z_I \\
     	0 \\
        -H\left(\bm{x}_I\right) x_I \\
  \end{array}\right], 
  \bm{u}^h\left(\bm{x}_I\right) = \left[ \begin{array}{c}
    	-H\left(\bm{x}_I\right) y_I \\
     	H\left(\bm{x}_I\right) x_I \\
        0 \\
  \end{array}\right] 
\end{align}

\begin{figure}
\centering
	\begin{subfigure}[b]{0.2\textwidth}
	\centering
		\includegraphics[height=0.75\textwidth]{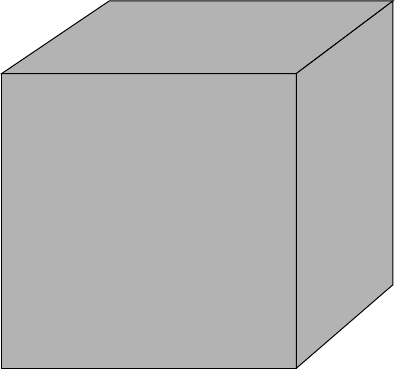}
		\caption{Original subdomain.}
	\end{subfigure}
	\begin{subfigure}[b]{0.4\textwidth}
	\centering
		\includegraphics[height=0.375\textwidth]{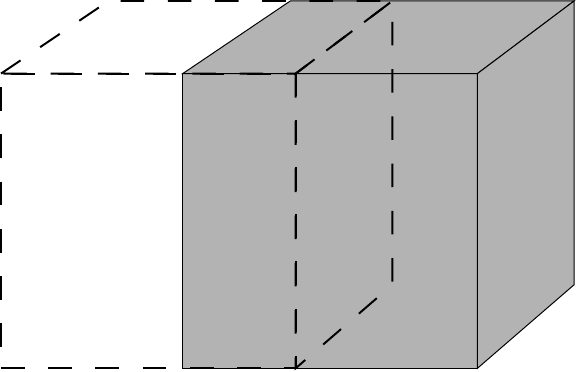}
		\caption{Rigid translation.}
	\end{subfigure}
	\begin{subfigure}[b]{0.2\textwidth}
	\centering
		\includegraphics[height=0.85\textwidth]{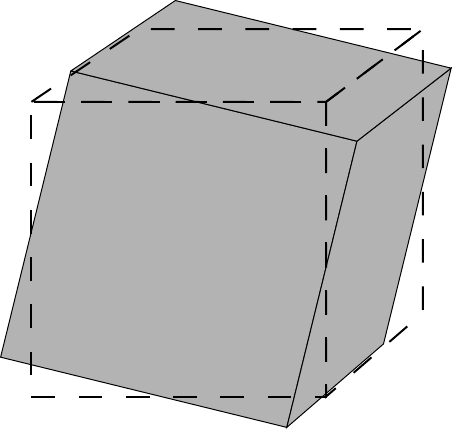}
		\caption{Rigid rotation.}
	\end{subfigure}
	\begin{subfigure}[b]{0.2\textwidth}
	\centering
		\includegraphics[height=0.75\textwidth]{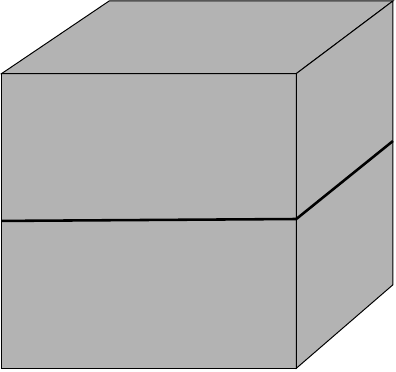}
		\caption{Cracked subdomain.}
	\end{subfigure}
	\begin{subfigure}[b]{0.4\textwidth}
	\centering
		\includegraphics[height=0.375\textwidth]{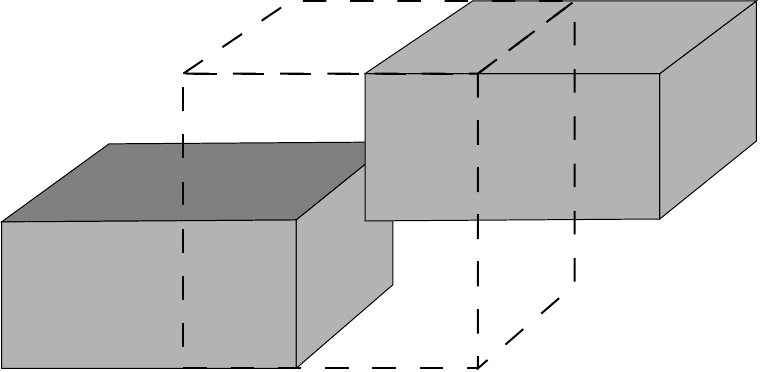}
		\caption{Rigid translation.}
	\end{subfigure}
	\begin{subfigure}[b]{0.2\textwidth}
	\centering
		\includegraphics[height=0.85\textwidth]{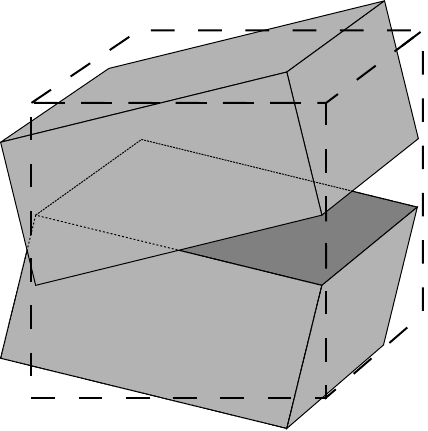}
		\caption{Rigid rotation.}
	\end{subfigure}
	
	\caption{Standard and enriched deflation modes for a subdomain.}
	\label{fig:subdomain_modes}
\end{figure}

\noindent These modes, in accordance to the modes of \autoref{eq:dcg_deflation_space} correspond to rigid translation and rotation of the two parts of the subdomain, lying above and below the crack, as illustrated in \autoref{fig:subdomain_modes}.

In order to derive the corresponding nodal values of the resulting deflation vectors, we start with the first mode considered, as well as the displacement approximation of \autoref{eq:xfem_displ}. Then, we observe that the following values for the standard and enriched degrees of freedom satisfy the first of \autoref{eq:enriched_deflated_modes}:

\begin{equation}
    \bm{u}_I = \left[ \begin{array}{c}
    	H\left(\bm{x}_I\right) \\
     	0 \\
        0 \\
    \end{array}\right],
    \bm{a}_I = \left[ \begin{array}{c}
    	1 \\
     	0 \\
        0 \\
    \end{array}\right]
\end{equation}

\noindent Similarly, the nodal values for the remaining modes can be derived, leading to the following deflation  matrices:

\begin{equation}
    \bm{W}^{\text{enr}_u}_I =\left[	
	\begin{array}{c c c c c c}
    	H\left(\bm{x}_I\right) & 0 & 0 &  0   &  H\left(\bm{x}_I\right) z_I & - H\left(\bm{x}_I\right)y_I  \\
     	0 & H\left(\bm{x}_I\right) & 0 & -H\left(\bm{x}_I\right) z_I &  0   &  H\left(\bm{x}_I\right) x_I  \\
        0 & 0 & H\left(\bm{x}_I\right) &  H\left(\bm{x}_I\right) y_I & -H\left(\bm{x}_I\right) x_I &  0
  \end{array}\right]
  \label{eq:dcg_deflation_space_enr_std}
\end{equation}

\begin{equation}
    \bm{W}^{\text{enr}_a}_I =\left[	
	\begin{array}{c c c c c c}
    	1 & 0 & 0 &  0   &  z_I & -y_I  \\
     	0 & 1 & 0 & -z_I &  0   &  x_I  \\
        0 & 0 & 1 &  y_I & -x_I &  0
  \end{array}\right]
  \label{eq:dcg_deflation_space_enr_enr}
\end{equation}

\noindent corresponding to the standard and enriched degrees of freedom respectively.

\subsection{Preconditioning and mesh partitioning}\label{preconditioning}


As shown in \autoref{deflation}, deflation can be combined with a preconditioner to further increase efficiency. Since lower eigenvalues of the system matrix are removed through deflation, the preconditioner should ideally remove higher eigenvalues to effectively reduce the condition number. If viewed as a multi-grid method, then the preconditioner assumes the role of a smoother, aiming at removing high frequency components of the error. Herein, this task is performed by a block-Jacobi preconditioner, which, additionally, can be computed completely in parallel. Block-Jacobi preconditioners approximate the system matrix, and subsequently its inverse, by a block diagonal matrix, obtained by removing off-diagonal blocks from the original matrix:

\begin{equation}
    \label{eq:block_jacobi}
	\bm{K} = \left[	
	\begin{array}{c c c c}
    	\bm{K_{11}}   & \bm{K_{12}}   & \cdots & \bm{K_{1n_b}} \\
     	\bm{K_{21}}   & \bm{K_{22}}   & \cdots & \bm{K_{2n_b}} \\
     	\vdots        & \vdots        & \ddots & \vdots \\
     	\bm{K_{n_b1}} & \bm{K_{n_b2}} & \cdots & \bm{K_{n_bn_b}}
  \end{array}\right]\approx \left[	
	\begin{array}{c c c c}
    	\bm{K_{11}} & \bm{0}      & \cdots & \bm{0} \\
     	\bm{0}      & \bm{K_{22}} & \cdots & \bm{0} \\
     	\vdots      & \vdots      & \ddots & \vdots \\
     	\bm{0}      & \bm{0}      & \cdots & \bm{K_{n_bn_b}}
  \end{array}\right]
\end{equation}

\noindent where blocks $\bm{K_{ij}}$ can be obtained by grouping together sets of unknowns. Since $\bm{K}$ is, in practice, sparse, the specific way in which unknowns are grouped can affect the performance of the preconditioner dramatically. More specifically, grouping dofs of neighbouring nodes together results in most of the off-diagonal blocks of $\bm{K}$ being zero, rendering the approximation of \autoref{eq:block_jacobi} more accurate. Therefore, groups can be selected to minimise the number of non-zero off-diagonal terms. To maximise the effectiveness the block Jacobi preconditioner, METIS~\cite{Karypis1997} is used to partition the mesh into subdomains, while minimising interactions between subdomains.

Regarding subdomains containing enriched nodes, two aternatives would be possible. The first would consist of constructing different blocks for the standard and enriched part of the stiffness matrix, and would resemble the approach used in some existing works dealing with preconditioning for XFEM/GFEM~\cite{Menk2011,Fillmore2017}. This approach would have the advantage of keeping the part of the preconditioner corresponding to the standard part of the approximation unaltered during a crack propagation simulation. The second alternative, which is chosen here, consists of modifying blocks of enriched subdomains to include enriched degrees of freedom, leading to the following structure:

\begin{equation}
    \label{eq:block_jacobi_enr}
	\bm{K} \approx \left[	
	\begin{array}{c c c c c c c c}
    	\bm{K_{11}} & \bm{0}      & \cdots & \bm{0} & \bm{0} & \cdots & \bm{0} \\
     	\bm{0}      & \bm{K_{22}} & \cdots & \bm{0}& \bm{0}  & \cdots &\bm{0} \\
     	\vdots      & \vdots      & \ddots & \vdots& \vdots  & \ddots & \vdots \\
     	\bm{0}      & \bm{0}      & \cdots & \bm{K_{FFii}} & \bm{K_{FXii}} & \cdots & \bm{0} \\
     	\bm{0}      & \bm{0}      & \cdots & \bm{K_{FXii}}^T & \bm{K_{XXii}} & \cdots & \bm{0} \\
     	\vdots      & \vdots      & \ddots & \vdots& \vdots  & \ddots & \vdots \\
     	\bm{0}      & \bm{0}      & \cdots &  \bm{0} & \bm{0}  & \cdots & \bm{K_{n_bn_b}}
  \end{array}\right]
\end{equation}

\noindent where $\bm{K_{FFii}}, \bm{K_{XXii}},\bm{K_{FXii}}$ are the standard, enriched and interaction parts of block $ii$ of the stiffness matrix. In the above, it is assumed that subdomain $i$, to which block $ii$ corresponds, contains enriched degrees of freedom, while subdomains 1,2 and $n_b$ do not.

The above choice can be justified by the fact that including enriched degrees of freedom in each block allows the preconditioner to remove linear dependencies between the standard and enriched part of the approximation, significantly decreasing the condition number of the initial matrix. Furthermore, the computational overhead for updating enriched blocks of the preconditioner in crack propagation simulations is small since, as also mentioned in \autoref{linear system of equations}, the number of enriched degrees of freedom, and as a result enriched subdomains, is only a small fraction of the total number of unknowns, and subdomains respectively.
\section{Numerical examples}\label{numerical examples}

\paragraph{Implementation}

The proposed approach is implemented in our in-house C++ code using the Eigen~\cite{eigen} library for linear algebra operations. The direct solutions employed in Algorithms \ref{alg:A-DEF2_u0} and \ref{alg:A-DEF2} are carried out with Pardiso \cite{Schenk2000,Schenk2002,Schenk2004}, a state of the art parallel direct solver from Intel, while the direct solutions required for the block Jacobi preconditioner are carried out using the Cholesky solver offered by Eigen~\cite{eigen}, which, for small systems, as the ones arising within the preconditioner, is more efficient than Pardiso. The only parts of the solver/preconditioner that are implemented in parallel, are the matrix-vector multiplications involved in the CG algorithm, for which the shared memory parallelisation offered by Eigen is employed, and the block Jacobi preconditioner, for which, OpenMP is employed to carry out factorisations and backward substitutions in parallel. The relative tolerance for the CG solver is set to $10^{-8}$.

All meshes are generated using gmsh~\cite{Geuzaine2009} and results are visualized using Paraview~\cite{Ahrens2005,Ayachit2015}. It should be noted that adopted meshes are not always the optimal choice for the problem tested, rather they are aimed at illustrating the performance of the proposed approach for systems of varying size. Optimised meshes can be obtained through local refinement at the vicinity of the crack or even through adaptive refinement based on error estimators \cite{Bordas2007b,Jin2017,Duflot2008}.

All of the following examples were ran on a workstation equipped with an Intel Xeon E3-1275 quad core processor, running at 3.80GHz, and 32GB of memory.

\paragraph{Crack propagation} Crack propagation lengths are computed using a Paris law, as described in Agathos et al.~\cite{Agathos2018-a} and crack propagation angles are computed using the maximum circumferential stress criterion. Stress Intensity Factors (SIFs) and energy release rates, are computed using the interaction integral method, as described in Agathos et al.~\cite{Agathos2016}.

\subsection{Notched beam under three point bending}

\begin{figure}
	\centering
		\includegraphics[width=0.6\textwidth]{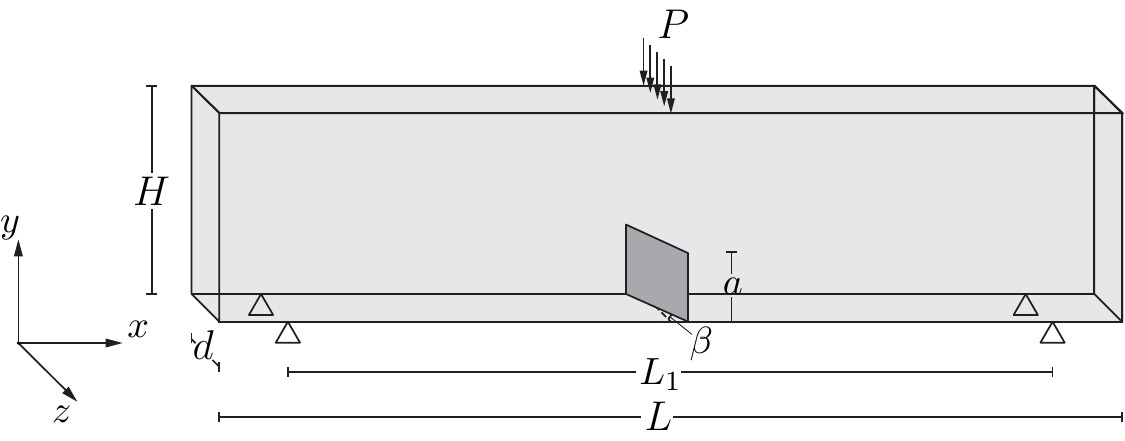}
	\caption{Notched beam under three point bending. The geometrical parameters of the problem are $L=260$ mm, $L_1=240$ mm, $H=60$ mm, $d=10$ mm, $\alpha=20$ mm, $\beta=45^{\circ}$.}
	\label{fig:beam_notch_inclined}
\end{figure}

The first example involves crack propagation in a notched beam under three point bending, which has already been the subject of numerical \cite{Sadeghirad2016,Agathos2017} and experimental studies \cite{Buchholz2004}. As illustrated in Figure~\ref{fig:beam_notch_inclined}, the notch is not normal to the axis of the beam, thus resulting in non-planar crack propagation. The geometrical parameters of the problem are $L=260$ mm, $L_1=240$ mm, $H=60$ mm, $d=10$ mm, $\alpha=20$ mm, $\beta=0^{\circ}, 45^{\circ}$, while the material parameters are $E=2.1 \times 10^5$ N/mm$^2$, $\nu=0.3$ and the applied load $P=2$ kN.

\begin{figure}
	\begin{subfigure}[b]{0.5\textwidth}
		\includegraphics[width=1.0\textwidth]{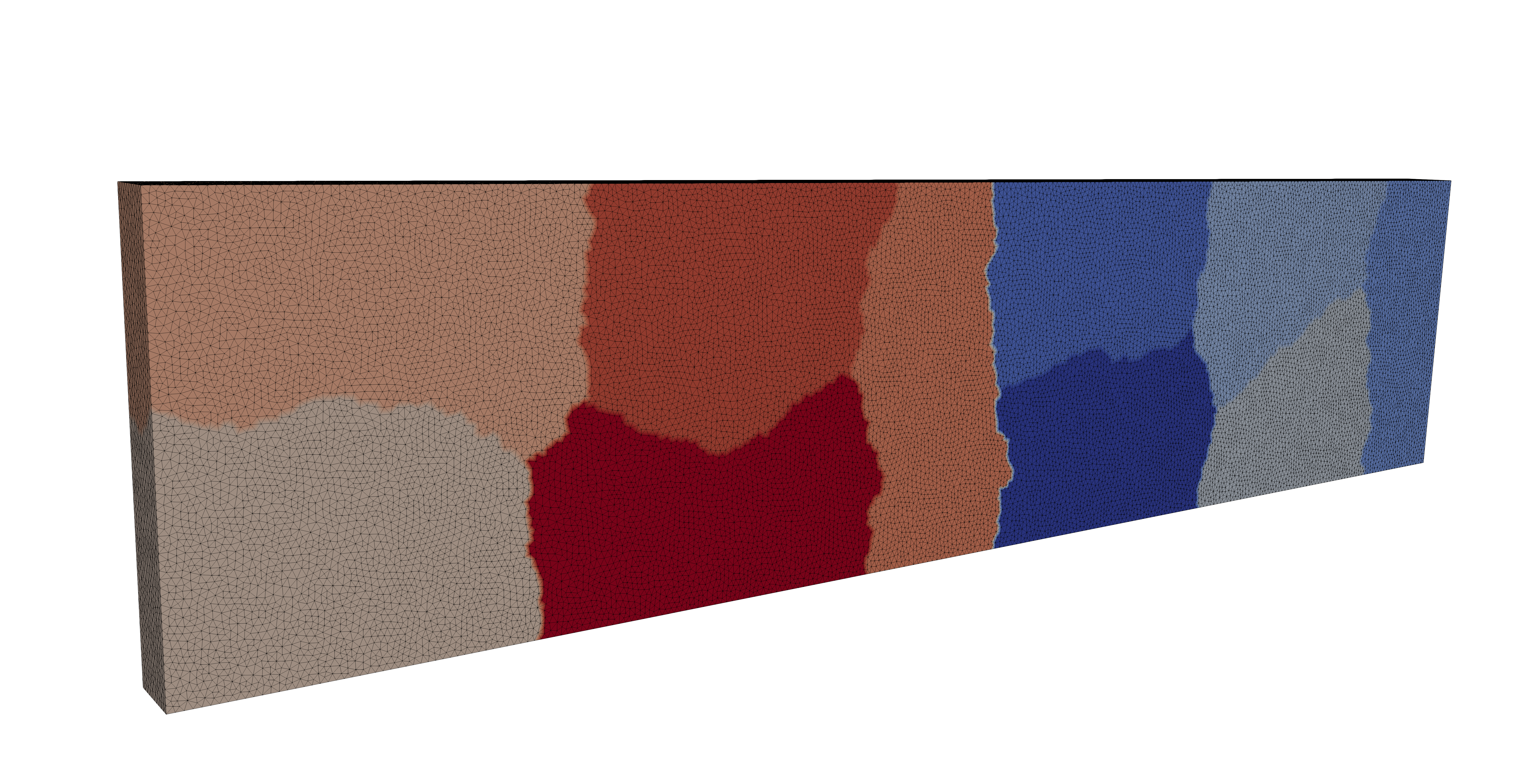}
		\caption{10 subdomains.}
	\end{subfigure}
	\begin{subfigure}[b]{0.5\textwidth}
		\includegraphics[width=1.0\textwidth]{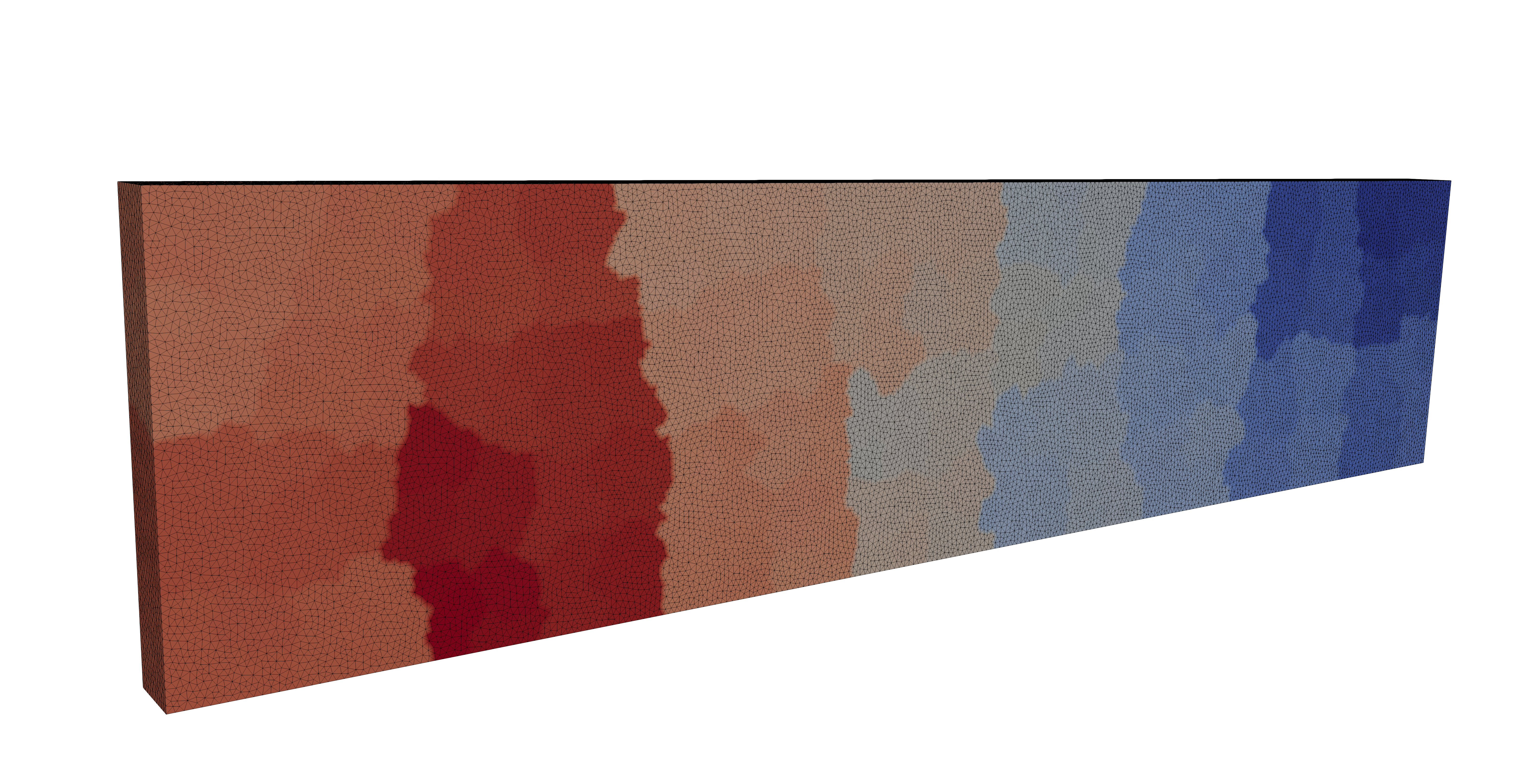}
		\caption{1,000 subdomains.}
	\end{subfigure}
	\caption{Notched beam under three point bending. Subdomains used for constructing the deflation space for two different cases for the mesh with $h=1$mm. Different colours correspond to different subdomains.}
	\label{fig:beam_subdomains}
\end{figure}

Three different levels of mesh refinement are used, the statistics of which are provided in \autoref{tab:beam_mesh_stats}. Moreover, in \autoref{fig:beam_subdomains}, the mesh with mesh size $h=1$mm is illustrated along with the corresponding subdomains for two different cases. For all cases, a geometrical enrichment strategy is used with an enrichment radius $r_e=3$mm.

\begin{table}
\centering
\caption{Notched beam under three point bending. Mesh statistics for three different refinement levels, corresponding to different mesh sizes $h$. The number of enriched dofs refers to the initial configuration, before any crack propagation occurs.}
\begin{tabular}{ |l| c| c| c|}
\hline
              &  $h=2$mm  &  $h=1$mm  &  $h=0.5$mm \\
\hline
elements      &  125,179  &  676,949  &  4,036,623 \\
nodes         &  26,405   &  134,808  &  753,639   \\
dofs          &  79,215   &  404,424  &  2,260,917 \\
enriched dofs &  846      &  5,535    &  26,550    \\
\hline
\end{tabular}
\label{tab:beam_mesh_stats}
\end{table}

\subsubsection{Effectiveness and robustness of the proposed preconditioning scheme and deflation space for enriched approximations}

\begin{figure}
	\begin{subfigure}[b]{0.5\textwidth}
		\begin{tikzpicture}
    \begin{axis}[
        xmin = 1,
        xmax = 10000,
        ymin = 20,
        ymax = 20000,
        xtick = {1,10,100,1000,10000},
        xticklabels = {$10^0$,$10^1$,$10^2$,$10^3$,$10^4$},
        ytick = {100,1000,10000},
        yticklabels = {$10^2$,$10^3$,$10^4$},
        xmode = log,
        ymode = log,
        xlabel = {number of subdomains},
        ylabel = {number of iterations},
        grid = both,
        xminorgrids=false,
        yminorgrids=false,
        width=1\textwidth,
        height=0.75\textwidth,
        xticklabel style = {font =\fontsize{\figureFontSize pt}{10pt}\selectfont},
        yticklabel style = {font=\fontsize{\figureFontSize pt}{10pt}\selectfont},
        xlabel style = {font =\fontsize{\figureFontSize pt}{\figureFontSize pt}\selectfont},
        ylabel style = {font=\fontsize{\figureFontSize pt}{\figureFontSize pt}\selectfont},
        legend columns=2,
        legend style={at={(0,-0.3)},anchor=north west,
        nodes={font=\fontsize{\figureFontSize pt}{\figureFontSize pt}\selectfont}}
        ]
    
    \addplot[color = blue,line width=0.75pt] table[x=sd, y=j, col sep=comma] {Tikz_figures/beam_sd_iterations.csv};
    
    \addplot[color = red, dashed,line width=0.75pt] table[x=sd, y=je, col sep=comma] {Tikz_figures/beam_sd_iterations.csv};
    
    \addplot[color = blue, dashdotted,line width=0.75pt] table[x=sd, y=bj, col sep=comma] {Tikz_figures/beam_sd_iterations.csv};
    
    \addplot[color = red, densely dotted,line width=0.75pt] table[x=sd, y=bje, col sep=comma] {Tikz_figures/beam_sd_iterations.csv};

   	\addlegendentry{Jacobi}
   	\addlegendentry{Jacobi + enrichment}
   	\addlegendentry{Block Jacobi}
   	\addlegendentry{Block Jacobi + enrichment}
   
    \end{axis}
\end{tikzpicture}
	\end{subfigure}
	\begin{subfigure}[b]{0.5\textwidth}
		\begin{tikzpicture}
    \begin{axis}[
        xmin = 1,
        xmax = 10000,
        ymin = 3,
        ymax = 300,
        xtick = {1,10,100,1000,10000},
        xticklabels = {$10^0$,$10^1$,$10^2$,$10^3$,$10^4$},
        ytick = {10,100},
        yticklabels = {$10^1$,$10^2$},
        xmode = log,
        ymode = log,
        xlabel = {number of subdomains},
        ylabel = {time (s)},
        grid = both,
        xminorgrids=false,
        yminorgrids=false,
        width=1\textwidth,
        height=0.75\textwidth,
        xticklabel style = {font =\fontsize{\figureFontSize pt}{10pt}\selectfont},
        yticklabel style = {font=\fontsize{\figureFontSize pt}{10pt}\selectfont},
        xlabel style = {font =\fontsize{\figureFontSize pt}{\figureFontSize pt}\selectfont},
        ylabel style = {font=\fontsize{\figureFontSize pt}{\figureFontSize pt}\selectfont},
        legend columns=2,
        legend style={at={(0,-0.22)},anchor=north west,
        nodes={font=\fontsize{\figureFontSize pt}{\figureFontSize pt}\selectfont}}
        ]
    
    \addplot[color = blue,line width=0.75pt] table[x=sd, y=j, col sep=comma] {Tikz_figures/beam_sd_time.csv};
    
    \addplot[color = red, dashed,line width=0.75pt] table[x=sd, y=je, col sep=comma] {Tikz_figures/beam_sd_time.csv};
    
    \addplot[color = blue, dashdotted,line width=0.75pt] table[x=sd, y=bj, col sep=comma] {Tikz_figures/beam_sd_time.csv};
    
    \addplot[color = red, densely dotted,line width=0.75pt] table[x=sd, y=bje, col sep=comma] {Tikz_figures/beam_sd_time.csv};

    \addplot[color = black, dashed, line width=0.75pt] coordinates{
		(1,8.891)
		(10000,8.891)
	};
    
   	\addlegendentry{Jacobi}
   	\addlegendentry{Jacobi + enrichment}
   	\addlegendentry{Block Jacobi}
   	\addlegendentry{Block Jacobi + enrichment}
   	\addlegendentry{Pardiso}
   
    \end{axis}
\end{tikzpicture}
	\end{subfigure}
	\caption{Notched beam under three point bending. Performance of deflation in conjunction to a Jacobi and the proposed block Jacobi preconditioner with the standard and enriched deflation space.}
	\label{fig:beam_subdomain_performance}
\end{figure}
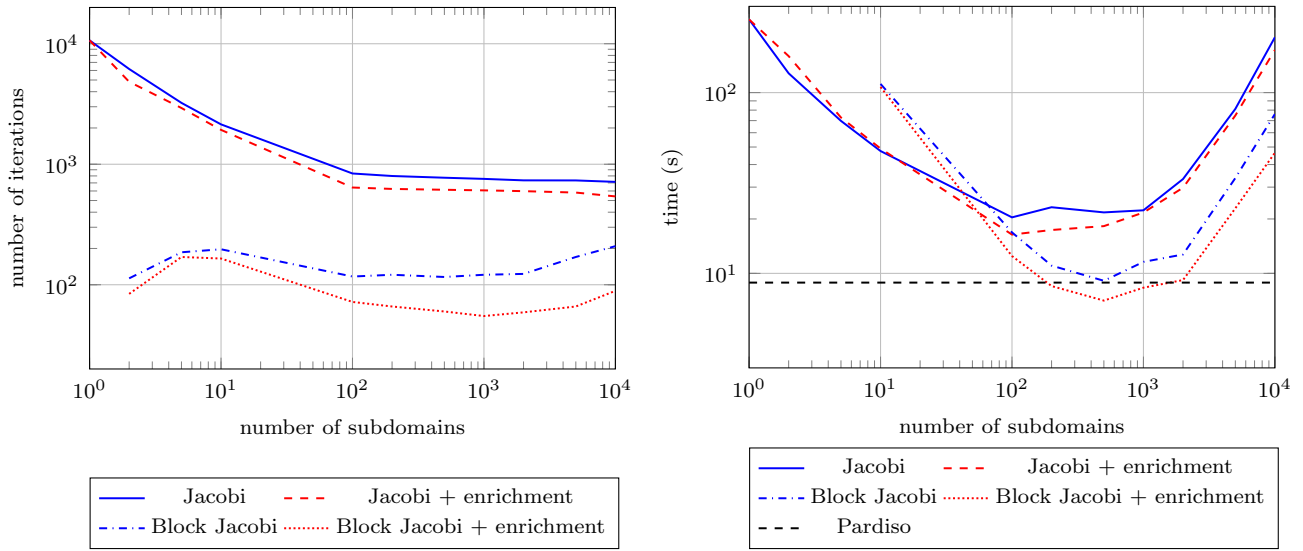

As a first test, in Figures~\ref{fig:beam_subdomain_performance_std} and \ref{fig:beam_subdomain_performance}, the effectiveness of both the proposed preconditioner and enriched deflation space are assessed in terms of numbers or iterations and wall time for the second level of mesh refinement ($h=1$mm) used and different numbers of subdomains. More specifically, results of \autoref{fig:beam_subdomain_performance_std}, refer to a beam without a notch, while in \autoref{fig:beam_subdomain_performance} the proposed approach is tested for the beam with the notch depicted in \autoref{fig:beam_notch_inclined}, which introduces enriched degrees of freedom. As a baseline for the comparisons, results were also obtained using a Jacobi preconditioner and Pardiso, which might often be the preferred option for problems of this size. For the Jacobi preconditioner, the single subdomain case corresponds to no deflation, while for the block Jacobi preconditioner, this case was not included since it would correspond to a solution with a direct solver.

Regarding the numbers of iterations, the Jacobi preconditioner results in the typical behaviour observed for deflated CG solvers, with iterations decreasing as the number of subdomains increases. However, it should be noted that, for the enriched case (\autoref{fig:beam_subdomain_performance}), this decrease is not as substantial as a result of linear dependences introduced by enrichment. The block Jacobi preconditioner results in a significant reduction in the number of iterations, especially in the enriched case, since it is much more effective in removing high frequency components of the error in each subdomain by accounting for interactions between different dofs. Moreover, the dependence of the required iterations on the number of subdomains is less significant, but more complex, consisting of an initial increase (2-10 subdomains), followed by a decrease (10-1,000 subdomains) and then a second increase (1,000-10,000 subdomains). This can be attributed to the effect of the decreasing block size as the number of subdomains increase. More specifically, for small numbers of subdomains, blocks are large, and the preconditioner is a very accurate approximation of the system matrix, leading to a reduced number of iterations. As a limit case, if only one subdomain was to be used, the solver would converge in a single iteration. On the other hand, as the number of subdomains increases, blocks become smaller and the block preconditioner becomes less and less accurate, the limit case being the Jacobi preconditioner. This trend is confirmed in both \autoref{fig:beam_subdomain_performance} and \ref{fig:beam_subdomain_performance_std}, where the numbers of iterations for the two cases tend towards the same value as subdomains increase. Between the two aforementioned regimes, the effect of preconditioning is combined to the effect of deflation leading to a decrease in the number of iterations.

In terms of wall time, in all cases it initially decreases as the number of subdomains increase, but tends to increase again after a certain point, as a result of iterations becoming more expensive for very large numbers of subdomains and the number of required iterations increasing in the case of the block Jacobi preconditioner. For the block Jacobi preconditioner, the initially increased computational time is attributed largely to the high cost of factorising blocks of the system matrix and performing backward substitutions. In fact, the peak in performance occurs for a block size, for which the direct solver used for these factorisations is the most efficient. Furthermore, while in the standard FE case (no enrichment, \autoref{fig:beam_subdomain_performance_std}) both preconditioners offer comparable performance, in the enriched case (\autoref{fig:beam_subdomain_performance}) the block Jacobi preconditioner is significantly more efficient, since it requires fewer iterations. This reduction comes as a result of linear dependences between enriched and standard dofs being removed within each block.

The use of an enriched deflation space always provides an additional improvement in terms of both iterations and wall time. When used in combination to the block Jacobi preconditioner, this improvement is more substantial. In terms of computational time, when used with an optimal number of subdomains, the proposed approach can be more efficient even than the direct solver, while compared to the baseline Jacobi preconditioned CG (1 subdomain case), it can be over 35 times as fast.

\begin{figure}
	\begin{subfigure}[b]{0.5\textwidth}
		\begin{tikzpicture}
    \begin{axis}[
        xmin = 1,
        xmax = 10000,
        ymin = 20,
        ymax = 20000,
        xtick = {1,10,100,1000,10000},
        xticklabels = {$10^0$,$10^1$,$10^2$,$10^3$,$10^4$},
        ytick = {100,1000,10000},
        yticklabels = {$10^2$,$10^3$,$10^4$},
        xmode = log,
        ymode = log,
        xlabel = {number of subdomains},
        ylabel = {number of iterations},
        grid = both,
        xminorgrids=false,
        yminorgrids=false,
        width=1\textwidth,
        height=0.75\textwidth,
        xticklabel style = {font =\fontsize{\figureFontSize pt}{10pt}\selectfont},
        yticklabel style = {font=\fontsize{\figureFontSize pt}{10pt}\selectfont},
        xlabel style = {font =\fontsize{\figureFontSize pt}{\figureFontSize pt}\selectfont},
        ylabel style = {font=\fontsize{\figureFontSize pt}{\figureFontSize pt}\selectfont},
        legend columns=2,
        legend style={at={(0,-0.3)},anchor=north west,
        nodes={font=\fontsize{\figureFontSize pt}{\figureFontSize pt}\selectfont}}
        ]
    
    \addplot[color = blue,line width=0.75pt] table[x=sd, y=j, col sep=comma] {Tikz_figures/beam_sd_iterations_std.csv};
    
    \addplot[color = blue, dashdotted,line width=0.75pt] table[x=sd, y=bj, col sep=comma] {Tikz_figures/beam_sd_iterations_std.csv};

   	\addlegendentry{Jacobi}
   	\addlegendentry{Block Jacobi}
   
    \end{axis}
\end{tikzpicture}
	\end{subfigure}
	\begin{subfigure}[b]{0.5\textwidth}
		\begin{tikzpicture}
    \begin{axis}[
        xmin = 1,
        xmax = 10000,
        ymin = 3,
        ymax = 300,
        xtick = {1,10,100,1000,10000},
        xticklabels = {$10^0$,$10^1$,$10^2$,$10^3$,$10^4$},
        ytick = {10,100},
        yticklabels = {$10^1$,$10^2$},
        xmode = log,
        ymode = log,
        xlabel = {number of subdomains},
        ylabel = {time (s)},
        grid = both,
        xminorgrids=false,
        yminorgrids=false,
        width=1\textwidth,
        height=0.75\textwidth,
        xticklabel style = {font =\fontsize{\figureFontSize pt}{10pt}\selectfont},
        yticklabel style = {font=\fontsize{\figureFontSize pt}{10pt}\selectfont},
        xlabel style = {font =\fontsize{\figureFontSize pt}{\figureFontSize pt}\selectfont},
        ylabel style = {font=\fontsize{\figureFontSize pt}{\figureFontSize pt}\selectfont},
        legend columns=2,
        legend style={at={(0,-0.22)},anchor=north west,
        nodes={font=\fontsize{\figureFontSize pt}{\figureFontSize pt}\selectfont}}
        ]
    
    \addplot[color = blue,line width=0.75pt] table[x=sd, y=j, col sep=comma] {Tikz_figures/beam_sd_time_std.csv};
    
    \addplot[color = blue, dashdotted,line width=0.75pt] table[x=sd, y=bj, col sep=comma] {Tikz_figures/beam_sd_time_std.csv};

    \addplot[color = black, dashed, line width=0.75pt] coordinates{
		(1,8.761)
		(10000,8.761)
	};
    
   	\addlegendentry{Jacobi}
   	\addlegendentry{Block Jacobi}
   	\addlegendentry{Pardiso}
   
    \end{axis}
\end{tikzpicture}
	\end{subfigure}
	\caption{Notched beam under three point bending. Performance of deflation in conjunction to a Jacobi and a block Jacobi preconditioner for the beam without a notch.}
	\label{fig:beam_subdomain_performance_std}
\end{figure}
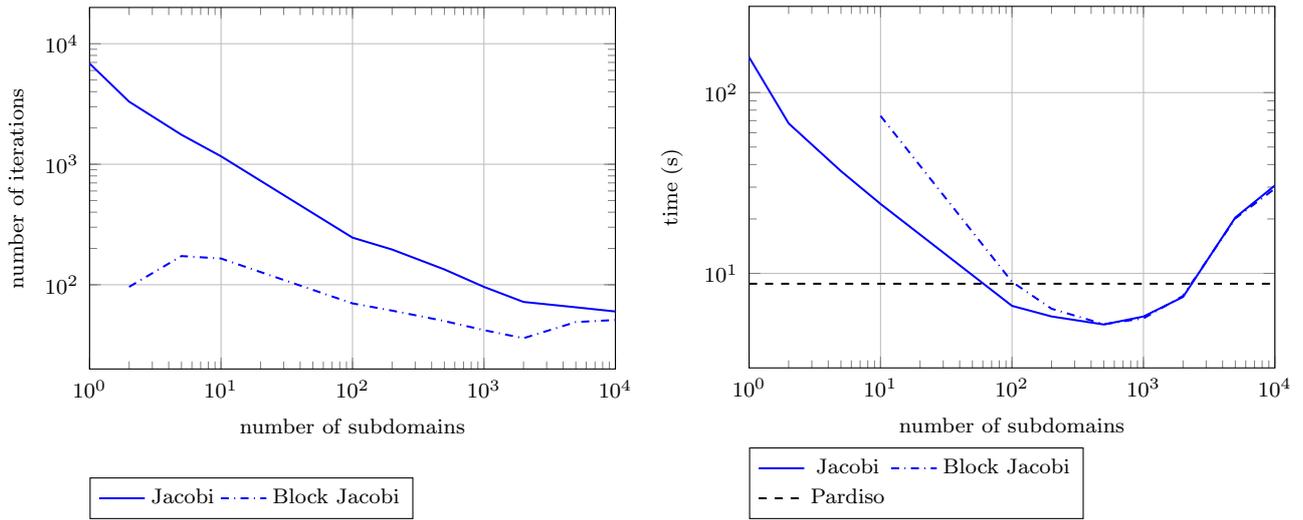

In \autoref{fig:beam_residual_iterations}, the relative residual of the solution is given for different numbers of CG iterations for all the alternatives considered and 500 subdomains. The proposed combination of enriched deflation vectors and block Jacobi preconditioning achieves the best performance, allowing to efficiently solve the problem even for very low tolerances.

\begin{figure}
    \centering
	\begin{tikzpicture}
    \begin{axis}[
        xmin = 1,
        xmax = 200,
        ymin = 1e-15,
        ymax = 4,
        xlabel = {iterations},
        ylabel = {relative residual},
        ymode = log,
        grid = both,
        xminorgrids=false,
        yminorgrids=false,
        width=0.5\textwidth,
        height=0.4\textwidth,
        xticklabel style = {font =\fontsize{\figureFontSize pt}{10pt}\selectfont},
        yticklabel style = {font=\fontsize{\figureFontSize pt}{10pt}\selectfont},
        xlabel style = {font =\fontsize{\figureFontSize pt}{\figureFontSize pt}\selectfont},
        ylabel style = {font=\fontsize{\figureFontSize pt}{\figureFontSize pt}\selectfont},
        legend style={at={(1.05,0)},anchor=south west, nodes={font=\fontsize{\figureFontSize pt}{\figureFontSize pt}\selectfont}}
        ]
    
    \addplot[color = blue,line width=0.75pt] table[x=i, y=r, col sep=comma] {Tikz_figures/DCG_residuals_J_500.csv};
    
    \addplot[color = red, dashed,line width=0.75pt] table[x=i, y=r, col sep=comma] {Tikz_figures/DCG_residuals_J_E_500.csv};
    
    \addplot[color = blue, dashdotted,line width=0.75pt] table[x=i, y=r, col sep=comma] {Tikz_figures/DCG_residuals_BJ_500.csv};
    
    \addplot[color = red, densely dotted,line width=0.75pt] table[x=i, y=r, col sep=comma] {Tikz_figures/DCG_residuals_BJ_E_500.csv};

   	\addlegendentry{Jacobi}
   	\addlegendentry{Jacobi + enrichment}
   	\addlegendentry{Block Jacobi}
   	\addlegendentry{Block Jacobi + enrichment}
   
    \end{axis}
\end{tikzpicture}
	\caption{Notched beam under three point bending. Relative residual of the CG solver for different numbers of iterations for all considered cases and 500 subdomains.}
	\label{fig:beam_residual_iterations}
\end{figure}
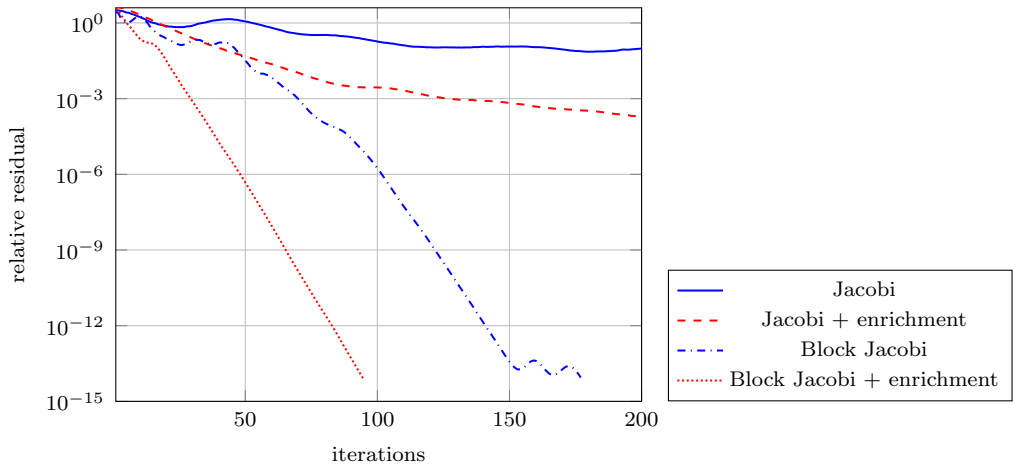

\begin{figure}
    \centering
	\begin{tikzpicture}
    \begin{axis}[
        xmin = 1,
        xmax = 4,
        ymin = 5,
        ymax = 16,
        xtick = {1,2,3,4},
        xlabel = {cpus},
        ylabel = {time (s)},
        grid = both,
        xminorgrids=false,
        yminorgrids=false,
        width=0.5\textwidth,
        height=0.4\textwidth,
        xticklabel style = {font =\fontsize{\figureFontSize pt}{10pt}\selectfont},
        yticklabel style = {font=\fontsize{\figureFontSize pt}{10pt}\selectfont},
        xlabel style = {font =\fontsize{\figureFontSize pt}{\figureFontSize pt}\selectfont},
        ylabel style = {font=\fontsize{\figureFontSize pt}{\figureFontSize pt}\selectfont},
        legend style={at={(1.05,0)},anchor=south west, nodes={font=\fontsize{\figureFontSize pt}{\figureFontSize pt}\selectfont}}
        ]
    
    \addplot[color = black,line width=0.75pt] table[x=c, y=p, col sep=comma] {Tikz_figures/beam_cpus_time.csv};
    
    \addplot[color = blue, dashed,line width=0.75pt] table[x=c, y=d500, col sep=comma] {Tikz_figures/beam_cpus_time.csv};
    

   	\addlegendentry{Pardiso}
   	\addlegendentry{PDCG-500 subdomains}
   
    \end{axis}
\end{tikzpicture}
	\caption{Notched beam under three point bending. Wall time for different number of cpu threads for Pardiso and the proposed approach using 500 subdomains.}
	\label{fig:beam_cpus_time}
\end{figure}
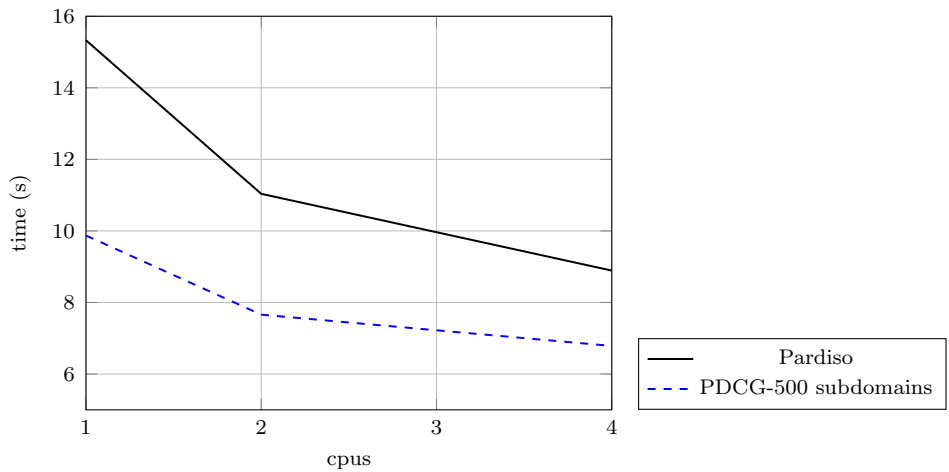

Although our current implementation is not optimised for parallel performance and scalability, in \autoref{fig:beam_cpus_time} a comparison is performed between the proposed approach and Pardiso, in terms of wall time, for different numbers of CPU threads. For the proposed approach 500 subdomains were used, which were found to yield the best performance. As can be seen, the proposed approach is more efficient, but it does not scale as well as Pardiso. This can be attributed to the fact that there are several parts of the solver, for instance mesh partitioning, that are not implemented in parallel, but are still accounted for in the computational time to provide a fair comparison.

\subsubsection{Performance of the proposed scheme for different levels of mesh refinement}

\begin{figure}
    \centering
	\begin{tikzpicture}
    \begin{axis}[
        xmin = 79215,
        xmax = 2260917,
        ymin = 40,
        ymax = 100,
        xlabel = {$n$},
        ylabel = {iterations},
        xmode = log,
        grid = both,
        xminorgrids=false,
        yminorgrids=false,
        width=0.5\textwidth,
        height=0.4\textwidth,
        xticklabel style = {font =\fontsize{\figureFontSize pt}{10pt}\selectfont},
        yticklabel style = {font=\fontsize{\figureFontSize pt}{10pt}\selectfont},
        xlabel style = {font =\fontsize{\figureFontSize pt}{\figureFontSize pt}\selectfont},
        ylabel style = {font=\fontsize{\figureFontSize pt}{\figureFontSize pt}\selectfont},
        legend style={at={(1.05,0)},anchor=south west, nodes={font=\fontsize{\figureFontSize pt}{\figureFontSize pt}\selectfont}}
        ]
    
    \addplot+[line width=0.75pt] table[x=n, y=i500, col sep=comma] {Tikz_figures/beam_dofs_iterations.csv};
    
    \addplot+[dashed,line width=0.75pt] table[x=n, y=i1000, col sep=comma] {Tikz_figures/beam_dofs_iterations.csv};
    
    \addplot+[dashdotted,line width=0.75pt] table[x=n, y=i2000, col sep=comma] {Tikz_figures/beam_dofs_iterations.csv};
    
    \addplot+[densely dotted,line width=0.75pt] table[x=n, y=i4000, col sep=comma] {Tikz_figures/beam_dofs_iterations.csv};
    
    \addlegendentry{$n/n_{sd}\approx500$}
    \addlegendentry{$n/n_{sd}\approx1,000$}
    \addlegendentry{$n/n_{sd}\approx2,000$}
    \addlegendentry{$n/n_{sd}\approx4,000$}
   
    \end{axis}
\end{tikzpicture}
	\caption{Notched beam under three point bending. Number of iterations required using the proposed preconditioner for systems of different size and different numbers of subdomains. For each of the lines the ratio between the toal number of unknowns and the number of subdomains ($n/n_{sd}$) is kept roughly constant, resulting in subdomains of similar size.}
	\label{fig:beam_dofs_iterations}
\end{figure}
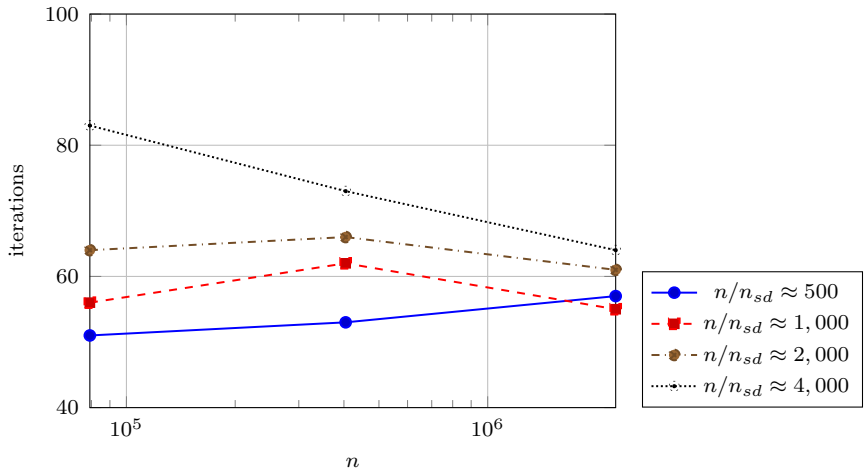

Next, the performance of the proposed scheme is tested for different mesh densities to provide some insight regarding the scalability of the approach. The results are given in \autoref{fig:beam_dofs_iterations} in terms of iterations for different numbers of unknowns. Each point in the figure corresponds to one of the meshes of \autoref{tab:beam_mesh_stats}, while different lines correspond to different ratios between the total number of unknowns ($n$) and the number of subdomains ($n_{sd}$). Thus, for each line in \autoref{fig:beam_dofs_iterations} the number of dofs per subdomain is kept roughly constant for the different mesh densities. As can be seen, although there are some fluctuations, the number of iterations does not change significantly for different system sizes, indicating that the proposed approach is scalable.

\subsubsection{Performance of the proposed scheme for propagating cracks}

\begin{figure}
	\begin{subfigure}[b]{0.5\textwidth}
		\includegraphics[width=1.0\textwidth]{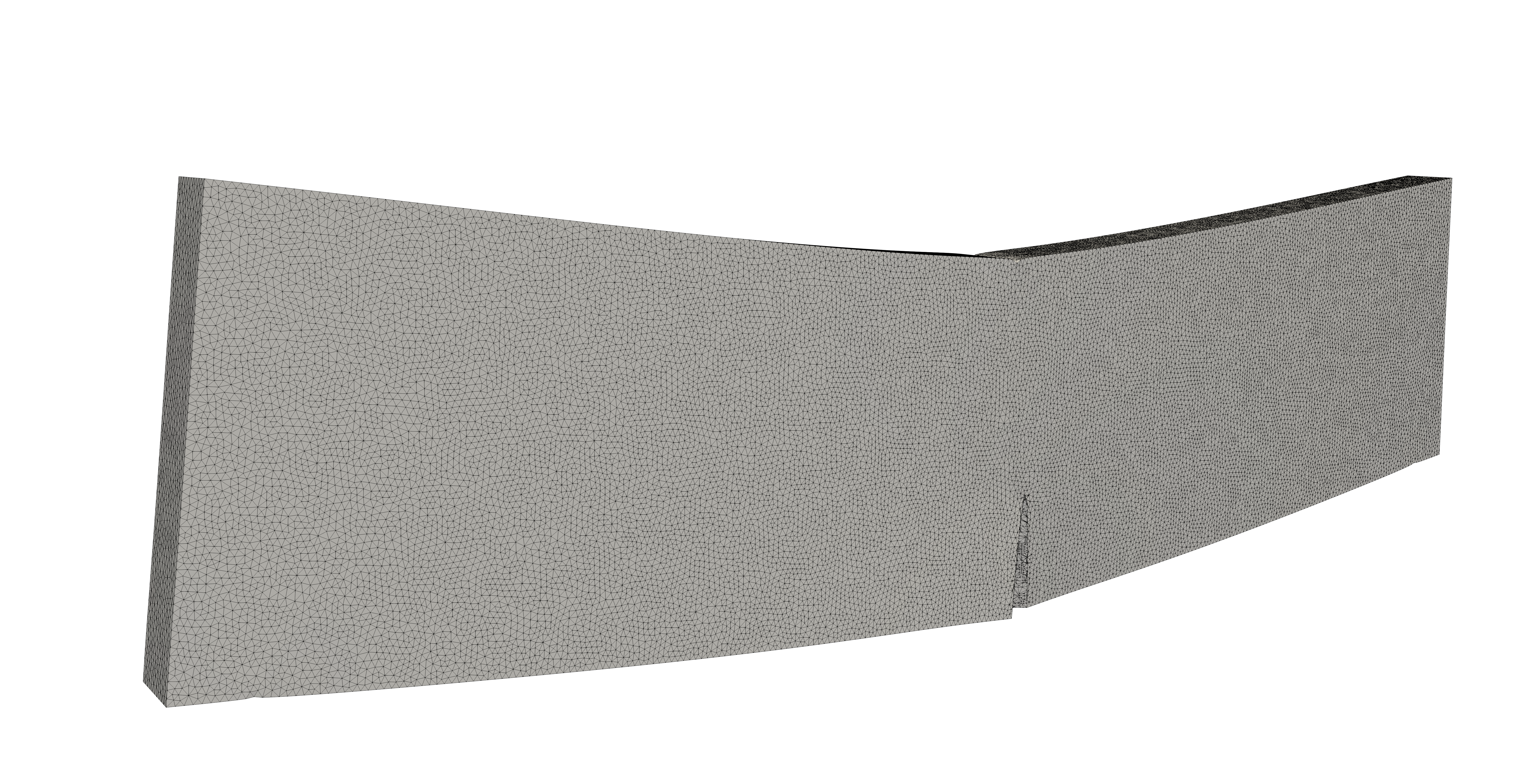}
		\caption{Initial crack.}
	\end{subfigure}
	\begin{subfigure}[b]{0.5\textwidth}
		\includegraphics[width=1.0\textwidth]{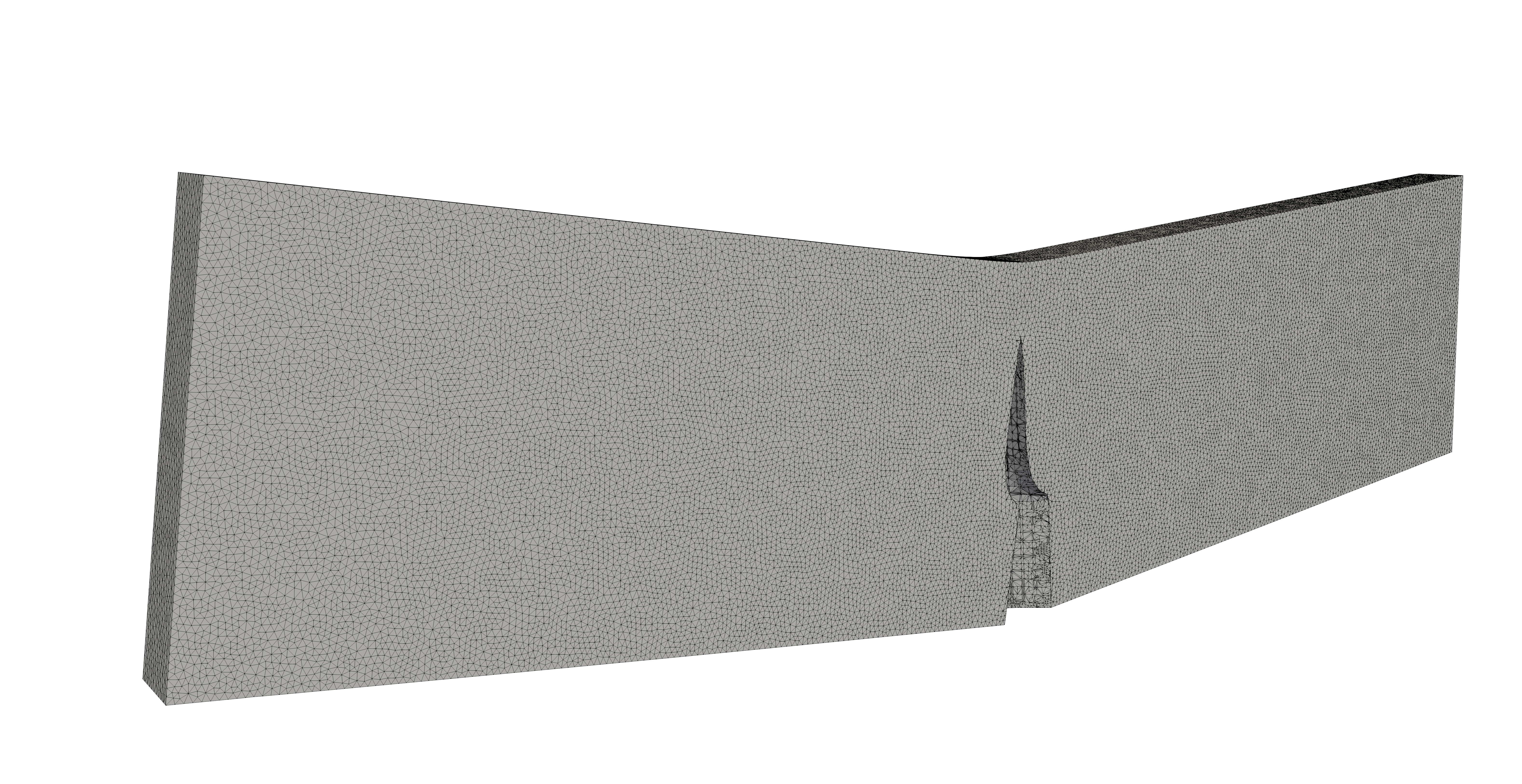}
		\caption{After 30 crack propagation steps.}
	\end{subfigure}
	\caption{Notched beam under three point bending. Deformed mesh for the initial and propagated crack after 30 steps of crack propagation for the mesh with $h=1$mm.}
	\label{fig:beam_deformed}
\end{figure}

\begin{figure}
    \centering
    \begin{subfigure}[t]{0.49\textwidth}
	    \begin{tikzpicture}
    \begin{axis}[
        xmin = 1,
        xmax = 30,
        ymin = 40,
        ymax = 1200,
        xlabel = {step},
        ylabel = {iterations},
        ymode = log,
        grid = both,
        xminorgrids=false,
        yminorgrids=false,
        width=1\textwidth,
        height=0.75\textwidth,
        xticklabel style = {font =\fontsize{\figureFontSize pt}{10pt}\selectfont},
        yticklabel style = {font=\fontsize{\figureFontSize pt}{10pt}\selectfont},
        xlabel style = {font =\fontsize{\figureFontSize pt}{\figureFontSize pt}\selectfont},
        ylabel style = {font=\fontsize{\figureFontSize pt}{\figureFontSize pt}\selectfont},
        legend columns=2,
        legend style={at={(0,-0.275)},anchor=north west,
        nodes={font=\fontsize{\figureFontSize pt}{\figureFontSize pt}\selectfont}}
        ]
    
    \addplot[color = blue, line width=0.75pt] table[x=stp, y=it_j, col sep=comma] {Tikz_figures/beam_steps_iterations_comparison.csv};
    
    \addplot[color = red, dashed, line width=0.75pt] table[x=stp, y=it_je, col sep=comma] {Tikz_figures/beam_steps_iterations_comparison.csv};
    
    \addplot[color = blue, dashdotted, mark = *, line width=0.75pt] table[x=stp, y=it_bj, col sep=comma] {Tikz_figures/beam_steps_iterations_comparison.csv};
    
    \addplot[color = red, densely dotted, mark = square*, line width=0.75pt] table[x=stp, y=it_bje, col sep=comma] {Tikz_figures/beam_steps_iterations_comparison.csv};
    
    \addlegendentry{Jacobi}
    \addlegendentry{Jacobi enriched}
    \addlegendentry{Block Jacobi}
    \addlegendentry{Block Jacobi enriched}
   
    \end{axis}

\end{tikzpicture}
	\end{subfigure}
	\begin{subfigure}[t]{0.49\textwidth}
	    \begin{tikzpicture}
    \begin{axis}[
        xmin = 1,
        xmax = 30,
        ymin = 3,
        ymax = 30,
        ytick = {7.5,15,30},
        yticklabels = {$7.5$,$15$,$30$},
        xlabel = {step},
        ylabel = {time (s)},
        ymode = log,
        grid = both,
        xminorgrids=false,
        yminorgrids=false,
        width=1\textwidth,
        height=0.75\textwidth,
        xticklabel style = {font =\fontsize{\figureFontSize pt}{10pt}\selectfont},
        yticklabel style = {font=\fontsize{\figureFontSize pt}{10pt}\selectfont},
        xlabel style = {font =\fontsize{\figureFontSize pt}{\figureFontSize pt}\selectfont},
        ylabel style = {font=\fontsize{\figureFontSize pt}{\figureFontSize pt}\selectfont},
        legend columns=2,
        legend style={at={(0,-0.2)},anchor=north west,
        nodes={font=\fontsize{\figureFontSize pt}{\figureFontSize pt}\selectfont}}
        ]
    
    \addplot[color = blue, line width=0.75pt] table[x=stp, y=t_j, col sep=comma] {Tikz_figures/beam_steps_time_comparison.csv};
    
    \addplot[color = red, dashed, line width=0.75pt] table[x=stp, y=t_je, col sep=comma] {Tikz_figures/beam_steps_time_comparison.csv};
    
    \addplot[color = blue, dashdotted, mark = *, line width=0.75pt] table[x=stp, y=t_bj, col sep=comma] {Tikz_figures/beam_steps_time_comparison.csv};
    
    \addplot[color = red, densely dotted, mark = square*, line width=0.75pt] table[x=stp, y=t_bje, col sep=comma] {Tikz_figures/beam_steps_time_comparison.csv};
    
    \addplot[color = black, dashed, line width=0.75pt] table[x=stp, y=p, col sep=comma] {Tikz_figures/beam_steps_time_comparison.csv};
    
    \addlegendentry{Jacobi}
    \addlegendentry{Jacobi enriched}
    \addlegendentry{Block Jacobi}
    \addlegendentry{Block Jacobi enriched}
    \addlegendentry{Pardiso}
   
    \end{axis}

\end{tikzpicture}
	\end{subfigure}
	
	\caption{Notched beam under three point bending. Number of iterations and wall time at each step of crack propagation for all alternatives considered using 500 subdomains.}
	\label{fig:beam_steps_iterations}
\end{figure}
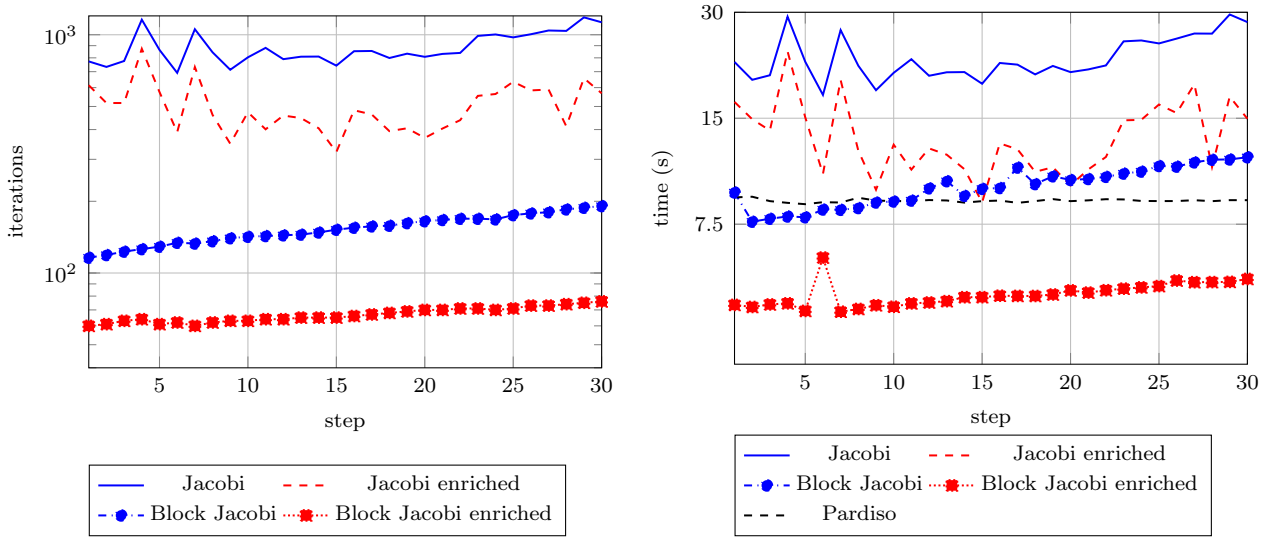

As a final test, the proposed approach is tested for a propagating crack. In \autoref{fig:beam_deformed}, the deformed mesh is illustrated for the initial crack and after 30 steps of crack propagation, while in \autoref{fig:beam_steps_iterations} the number of iterations and wall time required by the linear solver at each crack propagation step are given for the different alternatives considered using 500 subdomains. In general, similar trends to \autoref{fig:beam_subdomain_performance} can be observed, however, crack propagation tends to affect the preconditioners considered in different ways. In particular, in all cases, the number of iterations and wall time tend to increase as the crack propagates, which can be attributed to the increased number of degrees of freedom and its effect on the solution, which, as illustrated in \autoref{fig:beam_deformed}, can be quite substantial. The Jacobi preconditioner, exhibits a much more oscillatory behaviour, indicating that its performance is strongly affected by the way in which subdomains are intersected by the crack. While enriching the deflation space improves the situation in terms of overall performance, it does not lead to a less oscillatory behaviour. On the other hand, the block Jacobi preconditioner is much more robust with respect to crack propagation and when combined to an enriched deflation space it leads to a further improvement in performance.

It should be noticed that, the difference between the proposed approach and Pardiso in \autoref{fig:beam_steps_iterations} is more pronounced than in \autoref{fig:beam_subdomain_performance}, since the timings reported in \autoref{fig:beam_subdomain_performance} also include the time needed for domain partitioning, which in the crack propagation case only needs to be performed once. As an indication of the difference in performance, the total time required for linear solutions for this example using Pardiso is 262s, while using the proposed approach 144s, including the time required for partitioning and all other necessary tasks, such as computation of the standard and enriched deflation vectors, projection of the system matrix in the deflation space, initialisation and update of the preconditioner. Furthermore, this difference could have been even more pronounced if a less conservative tolerance was used for the iterative solver, which based on our experiments might have been possible without affecting the accuracy of the results.

\subsection{Notched bearing bracket}

\begin{figure}
	\centering
		\includegraphics[width=0.45\textwidth]{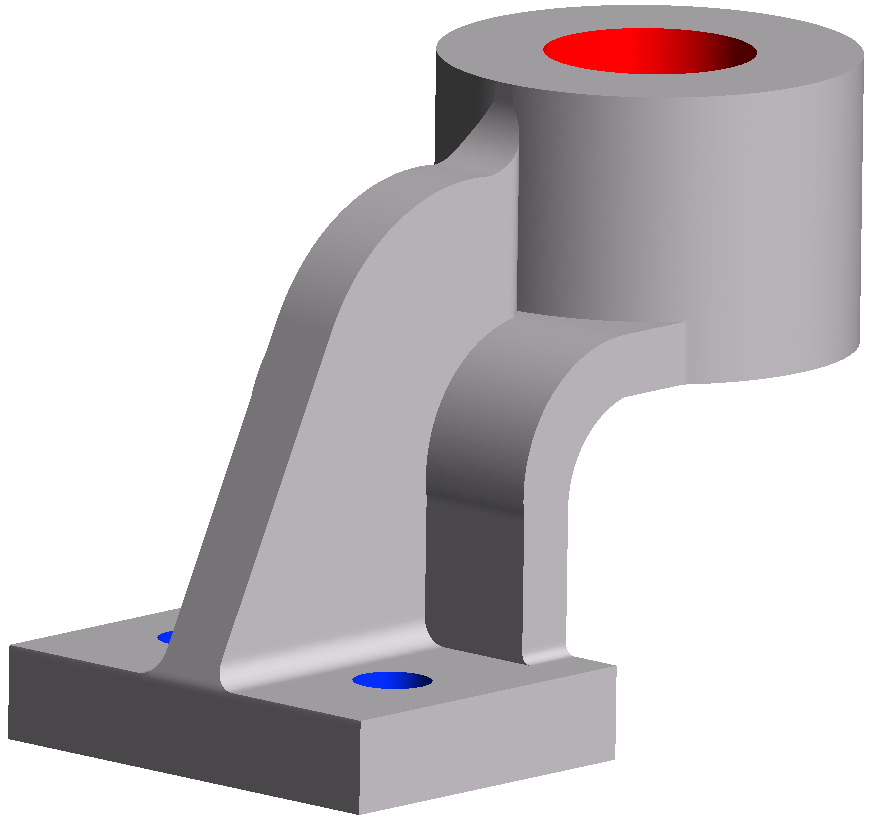}
	\caption{Notched bearing bracket geometry and boundary conditions.Blue colour corresponds to fixed displacements and red to a uniform traction in the vertical direction.}
	\label{fig:bracket}
\end{figure}

The next example involves the geometry of a bearing bracket, downloaded from \href{https://grabcad.com/library/bearing-bracket-26}{grabcad.com} and illustrated in \autoref{fig:bracket}. The same material properties as the previous example are used: $E=2.1 \times 10^5$ N/mm$^2$, $\nu=0.3$. Zero displacements are imposed at the two holes in the bottom (highlighted in blue) and uniform surface tractions are applied in the horizontal direction at the top hole (highlighted in red). The initial location of the notch is shown in \autoref{fig:bracket_initial_crack}.

An unstructured mesh with mesh size $h=1$mm, whose statistics are given in \autoref{tab:bracket_mesh_stats}, is used. For the enriched part, a geometrical enrichment strategy is used with an enrichment radius $r_e=3$mm.

\begin{table}
\centering
\caption{Notched bearing bracket. Mesh statistics for the mesh used. The number of enriched dofs refers to the initial configuration, before any crack propagation occurs.}
\begin{tabular}{ |l| c|}
\hline
              &  $h=1$mm  \\
\hline
elements      &   4,598,525 \\
nodes         &   793,975 \\
dofs          &   2,381,925 \\
enriched dofs &   15,552 \\
\hline
\end{tabular}
\label{tab:bracket_mesh_stats}
\end{table}

\subsubsection{Effectiveness and robustness of the proposed preconditioning scheme for enriched approximations}

For this example, only the enriched deflation space, combined with the Jacobi and block Jacobi preconditioners is considered. The standard deflation space is not considered, since in the previous example it was shown to be less effective than the enriched one, while the direct solver would be inefficient due to excessive memory requirements. 

\begin{figure}
	\begin{subfigure}[b]{0.5\textwidth}
		\begin{tikzpicture}
    \begin{axis}[
        xmin = 10,
        xmax = 10000,
        ymin = 50,
        ymax = 4000,
        xtick = {10,100,1000,10000},
        xticklabels = {$10^1$,$10^2$,$10^3$,$10^4$},
        ytick = {100,1000,10000},
        yticklabels = {$10^2$,$10^3$,$10^4$},
        xmode = log,
        ymode = log,
        xlabel = {number of subdomains},
        ylabel = {number of iterations},
        grid = both,
        xminorgrids=false,
        yminorgrids=false,
        width=1\textwidth,
        height=0.75\textwidth,
        xticklabel style = {font =\fontsize{\figureFontSize pt}{10pt}\selectfont},
        yticklabel style = {font=\fontsize{\figureFontSize pt}{10pt}\selectfont},
        xlabel style = {font =\fontsize{\figureFontSize pt}{\figureFontSize pt}\selectfont},
        ylabel style = {font=\fontsize{\figureFontSize pt}{\figureFontSize pt}\selectfont},
        legend columns=2,
        legend style={at={(0,-0.22)},anchor=north west,
        nodes={font=\fontsize{\figureFontSize pt}{\figureFontSize pt}\selectfont}}
        ]
    
    \addplot[color = blue,line width=0.75pt] table[x=sd, y=je, col sep=comma] {Tikz_figures/bracket_subdomain_iterations_time_std.csv};
    
    \addplot[color = blue, dashdotted,line width=0.75pt] table[x=sd, y=bje, col sep=comma] {Tikz_figures/bracket_subdomain_iterations_time_std.csv};

   	\addlegendentry{Jacobi}
   	\addlegendentry{Block Jacobi}
   
    \end{axis}
\end{tikzpicture}
	\end{subfigure}
	\begin{subfigure}[b]{0.5\textwidth}
		\begin{tikzpicture}
    \begin{axis}[
        xmin = 10,
        xmax = 10000,
        ymin = 50,
        ymax = 1500,
        xtick = {10,100,1000,10000},
        xticklabels = {$10^1$,$10^2$,$10^3$,$10^4$},
        ytick = {10,100,1000},
        yticklabels = {$10^1$,$10^2$,$10^3$},
        xmode = log,
        ymode = log,
        xlabel = {number of subdomains},
        ylabel = {time (s)},
        grid = both,
        xminorgrids=false,
        yminorgrids=false,
        width=1\textwidth,
        height=0.75\textwidth,
        xticklabel style = {font =\fontsize{\figureFontSize pt}{10pt}\selectfont},
        yticklabel style = {font=\fontsize{\figureFontSize pt}{10pt}\selectfont},
        xlabel style = {font =\fontsize{\figureFontSize pt}{\figureFontSize pt}\selectfont},
        ylabel style = {font=\fontsize{\figureFontSize pt}{\figureFontSize pt}\selectfont},
        legend columns=2,
        legend style={at={(0,-0.22)},anchor=north west,
        nodes={font=\fontsize{\figureFontSize pt}{\figureFontSize pt}\selectfont}}
        ]
    
    \addplot[color = blue,line width=0.75pt] table[x=sd, y=jet, col sep=comma] {Tikz_figures/bracket_subdomain_iterations_time_std.csv};
    
    \addplot[color = blue, dashdotted,line width=0.75pt] table[x=sd, y=bjet, col sep=comma] {Tikz_figures/bracket_subdomain_iterations_time_std.csv};
    
   	\addlegendentry{Jacobi}
   	\addlegendentry{Block Jacobi}
   
    \end{axis}
\end{tikzpicture}
	\end{subfigure}
	\caption{Notched bearing bracket. Performance of deflation with an enriched deflation space in conjunction to a Jacobi and a block Jacobi preconditioner for the bracket without a notch.}
	\label{fig:bracket_subdomain_performance_std}
\end{figure}
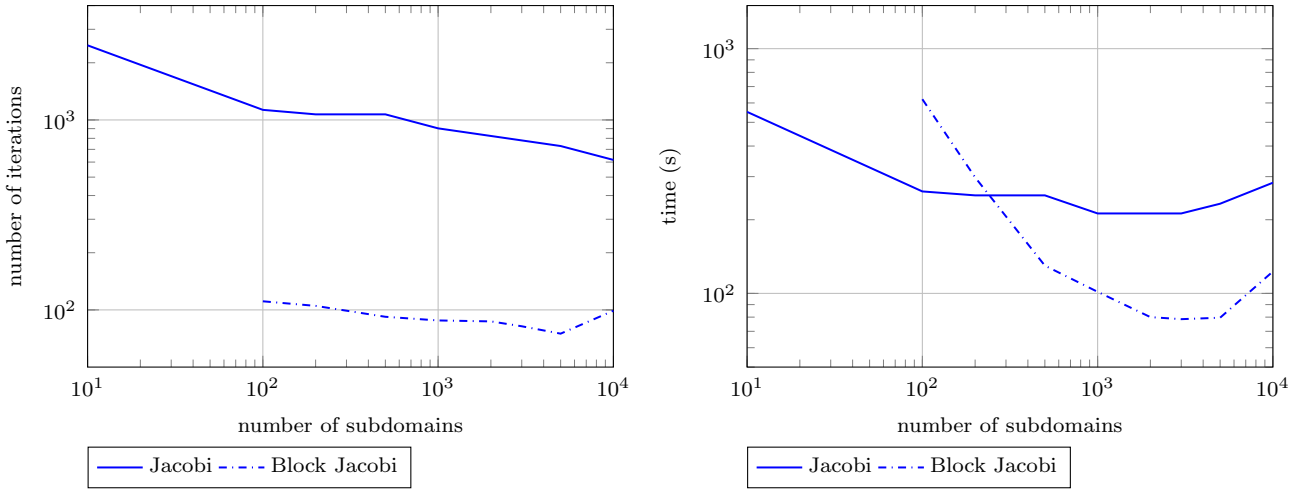

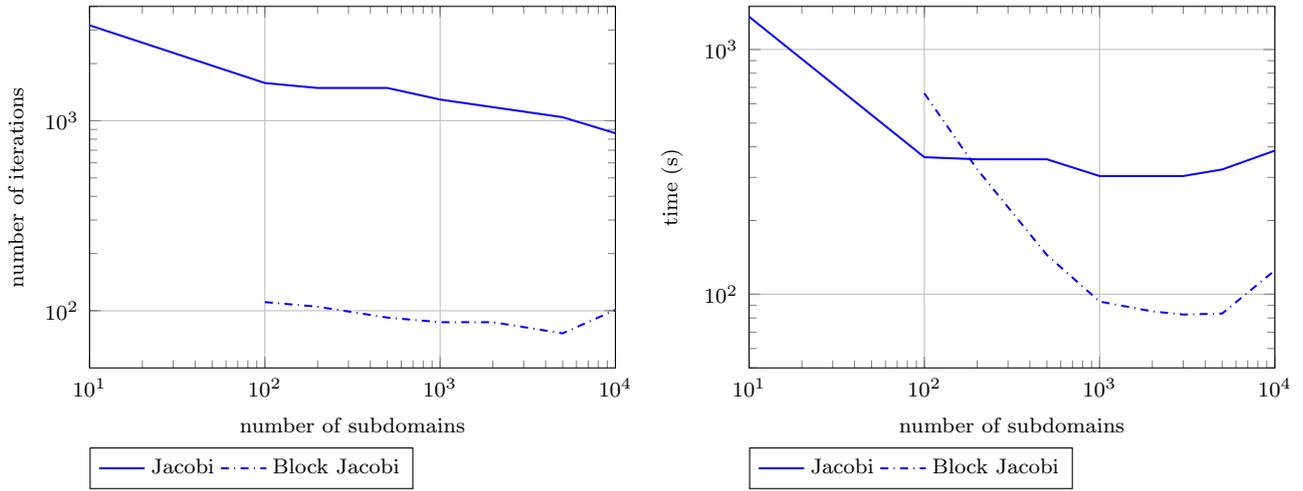
\begin{figure}
	\begin{subfigure}[b]{0.5\textwidth}
		\begin{tikzpicture}
    \begin{axis}[
        xmin = 10,
        xmax = 10000,
        ymin = 50,
        ymax = 4000,
        xtick = {10,100,1000,10000},
        xticklabels = {$10^1$,$10^2$,$10^3$,$10^4$},
        ytick = {100,1000,10000},
        yticklabels = {$10^2$,$10^3$,$10^4$},
        xmode = log,
        ymode = log,
        xlabel = {number of subdomains},
        ylabel = {number of iterations},
        grid = both,
        xminorgrids=false,
        yminorgrids=false,
        width=1\textwidth,
        height=0.75\textwidth,
        xticklabel style = {font =\fontsize{\figureFontSize pt}{10pt}\selectfont},
        yticklabel style = {font=\fontsize{\figureFontSize pt}{10pt}\selectfont},
        xlabel style = {font =\fontsize{\figureFontSize pt}{\figureFontSize pt}\selectfont},
        ylabel style = {font=\fontsize{\figureFontSize pt}{\figureFontSize pt}\selectfont},
        legend columns=2,
        legend style={at={(0,-0.22)},anchor=north west,
        nodes={font=\fontsize{\figureFontSize pt}{\figureFontSize pt}\selectfont}}
        ]
    
    \addplot[color = blue,line width=0.75pt] table[x=sd, y=je, col sep=comma] {Tikz_figures/bracket_subdomain_iterations_time.csv};
    
    \addplot[color = blue, dashdotted,line width=0.75pt] table[x=sd, y=bje, col sep=comma] {Tikz_figures/bracket_subdomain_iterations_time.csv};

   	\addlegendentry{Jacobi}
   	\addlegendentry{Block Jacobi}
   
    \end{axis}
\end{tikzpicture}
	\end{subfigure}
	\begin{subfigure}[b]{0.5\textwidth}
		\begin{tikzpicture}
    \begin{axis}[
        xmin = 10,
        xmax = 10000,
        ymin = 50,
        ymax = 1500,
        xtick = {10,100,1000,10000},
        xticklabels = {$10^1$,$10^2$,$10^3$,$10^4$},
        ytick = {10,100,1000},
        yticklabels = {$10^1$,$10^2$,$10^3$},
        xmode = log,
        ymode = log,
        xlabel = {number of subdomains},
        ylabel = {time (s)},
        grid = both,
        xminorgrids=false,
        yminorgrids=false,
        width=1\textwidth,
        height=0.75\textwidth,
        xticklabel style = {font =\fontsize{\figureFontSize pt}{10pt}\selectfont},
        yticklabel style = {font=\fontsize{\figureFontSize pt}{10pt}\selectfont},
        xlabel style = {font =\fontsize{\figureFontSize pt}{\figureFontSize pt}\selectfont},
        ylabel style = {font=\fontsize{\figureFontSize pt}{\figureFontSize pt}\selectfont},
        legend columns=2,
        legend style={at={(0,-0.22)},anchor=north west,
        nodes={font=\fontsize{\figureFontSize pt}{\figureFontSize pt}\selectfont}}
        ]
    
    \addplot[color = blue,line width=0.75pt] table[x=sd, y=jet, col sep=comma] {Tikz_figures/bracket_subdomain_iterations_time.csv};
    
    \addplot[color = blue, dashdotted,line width=0.75pt] table[x=sd, y=bjet, col sep=comma] {Tikz_figures/bracket_subdomain_iterations_time.csv};
    
   	\addlegendentry{Jacobi}
   	\addlegendentry{Block Jacobi}
   
    \end{axis}
\end{tikzpicture}
	\end{subfigure}
	\caption{Notched bearing bracket. Performance of deflation with an enriched deflation space in conjunction to a Jacobi and a block Jacobi preconditioner.}
	\label{fig:bracket_subdomain_performance}
\end{figure}

In Figures~\ref{fig:bracket_subdomain_performance_std} and \ref{fig:bracket_subdomain_performance}, the performance of the two preconditioners is illustrated for different numbers of subdomains in terms of iterations and computational time for the bracket without and with a notch respectively. In this example, due to the small size of the notch compared to the component, the response is only slightly affected, resulting in identical numbers of iterations for the healthy and cracked case and almost identical computational times. Moreover, the block Jacobi preconditioner reduces the number of required iterations by almost a full order of magnitude, compared to the Jacobi preconditioner. For higher numbers of subdomains, this also translates to a significant decrease in computational time.

\subsubsection{Performance of the proposed scheme for propagating cracks}

\begin{figure}
	\begin{subfigure}[b]{0.5\textwidth}
		\includegraphics[width=1.0\textwidth,trim={40cm 0 40cm 0},clip]{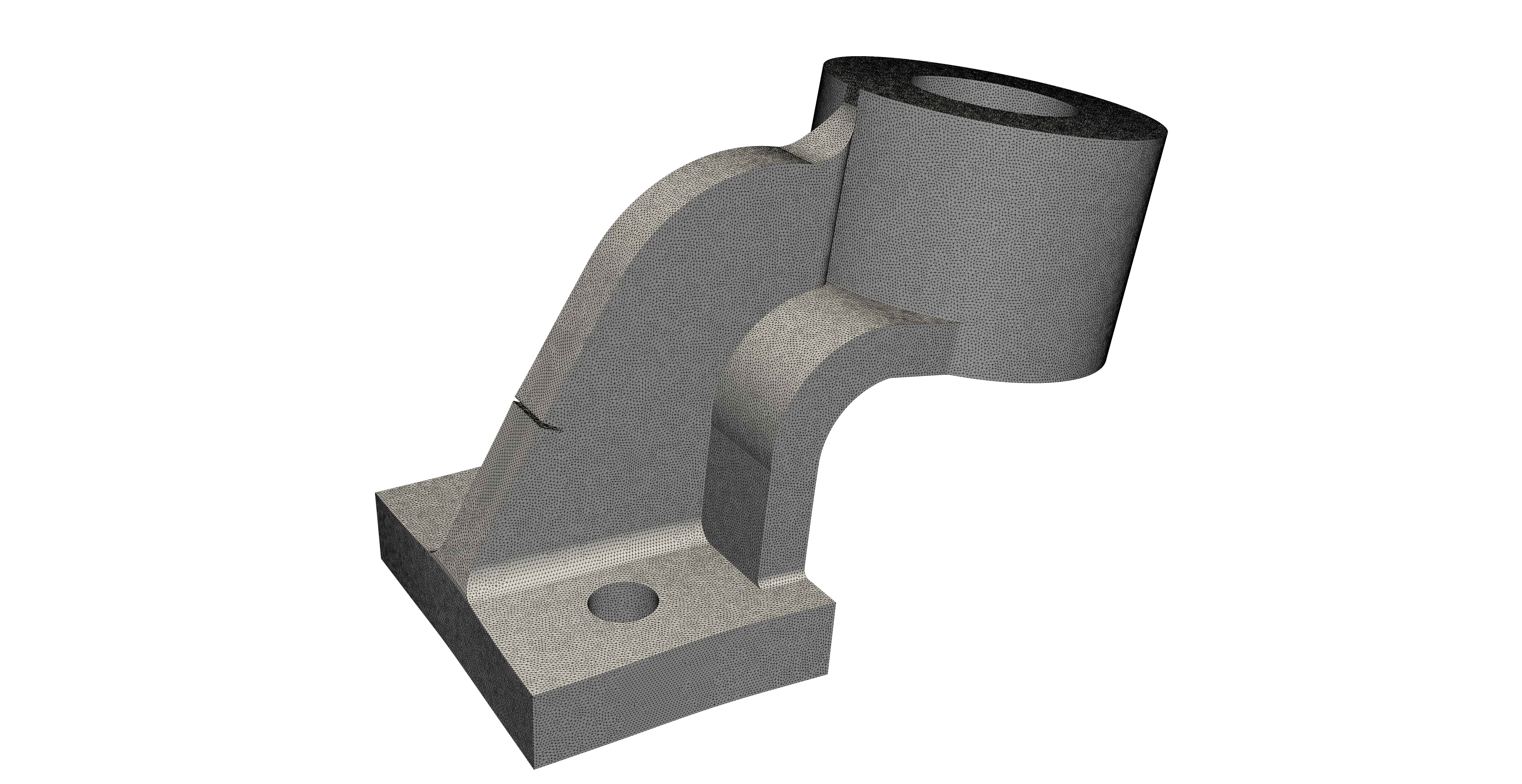}
		\caption{Initial crack.}\label{fig:bracket_initial_crack}
	\end{subfigure}
	\begin{subfigure}[b]{0.5\textwidth}
		\includegraphics[width=1.0\textwidth,trim={40cm 0 40cm 0},clip]{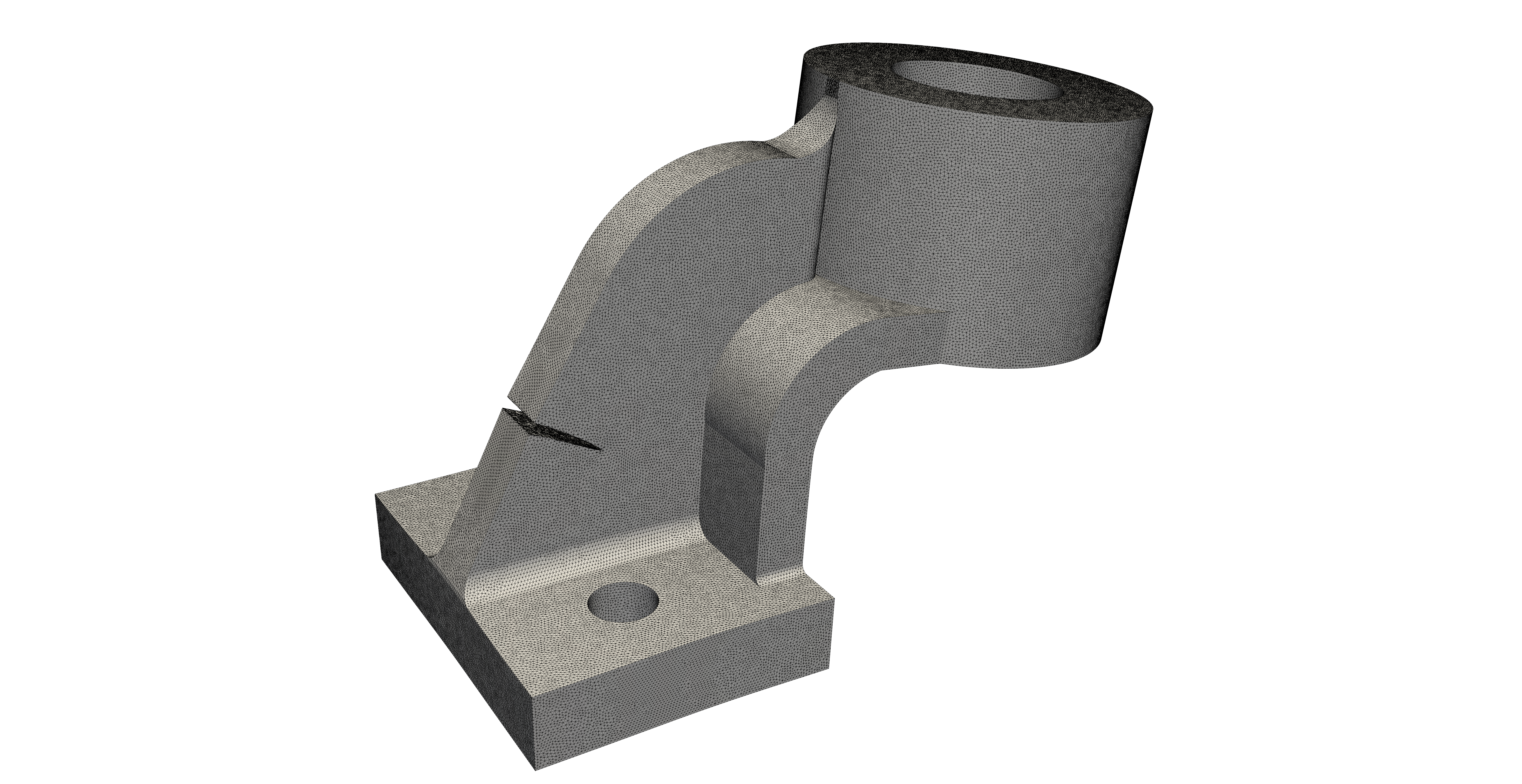}
		\caption{After 15 crack propagation steps.}\label{fig:bracket_final_crack}
	\end{subfigure}
	\caption{Notched bearing bracket. Deformed mesh for the initial and propagated crack after 15 steps of crack propagation.}
	\label{fig:bracket_cracks_deformed}
\end{figure}

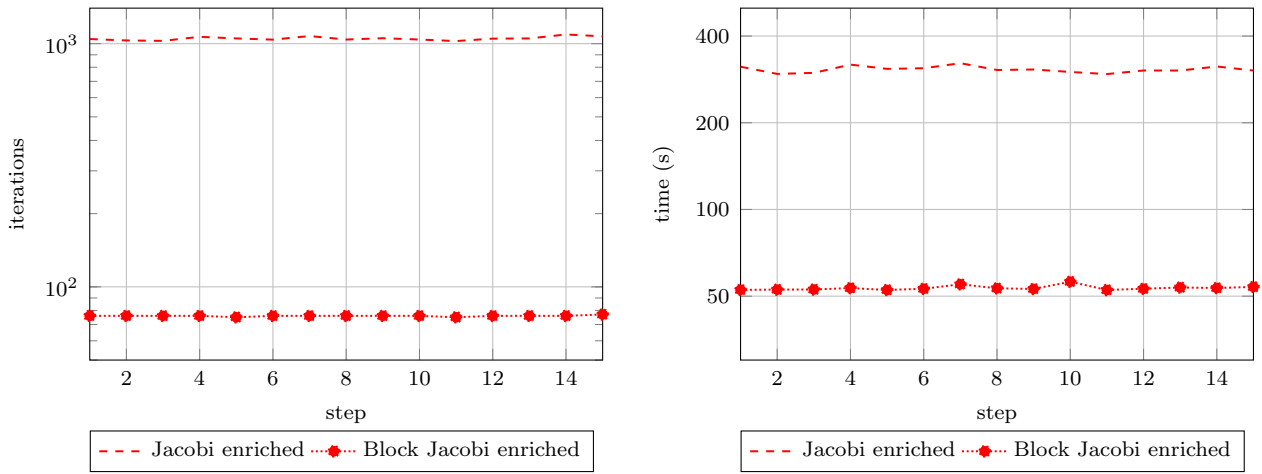
\begin{figure}
    \centering
    \begin{subfigure}[b]{0.49\textwidth}
	    \begin{tikzpicture}
    \begin{axis}[
        xmin = 1,
        xmax = 15,
        ymin = 50,
        ymax = 1400,
        xlabel = {step},
        ylabel = {iterations},
        ymode = log,
        grid = both,
        xminorgrids=false,
        yminorgrids=false,
        width=1\textwidth,
        height=0.75\textwidth,
        xticklabel style = {font =\fontsize{\figureFontSize pt}{10pt}\selectfont},
        yticklabel style = {font=\fontsize{\figureFontSize pt}{10pt}\selectfont},
        xlabel style = {font =\fontsize{\figureFontSize pt}{\figureFontSize pt}\selectfont},
        ylabel style = {font=\fontsize{\figureFontSize pt}{\figureFontSize pt}\selectfont},
        legend columns=2,
        legend style={at={(0,-0.2)},anchor=north west,
        nodes={font=\fontsize{\figureFontSize pt}{\figureFontSize pt}\selectfont}}
        ]
    
    \addplot[color = red, dashed, line width=0.75pt] table[x=stp, y=it_je, col sep=comma] {Tikz_figures/bracket_steps_iterations.csv};
    
    \addplot[color = red, densely dotted, mark = *, line width=0.75pt] table[x=stp, y=it_bje, col sep=comma] {Tikz_figures/bracket_steps_iterations.csv};
    
    \addlegendentry{Jacobi enriched}
    \addlegendentry{Block Jacobi enriched}
   
    \end{axis}

\end{tikzpicture}
	\end{subfigure}
	\begin{subfigure}[b]{0.49\textwidth}
	    \begin{tikzpicture}
    \begin{axis}[
        xmin = 1,
        xmax = 15,
        ymin = 30,
        ymax = 500,
        ytick = {50,100,200,400},
        yticklabels = {$50$,$100$,$200$,$400$},
        xlabel = {step},
        ylabel = {time (s)},
        ymode = log,
        grid = both,
        xminorgrids=false,
        yminorgrids=false,
        width=1\textwidth,
        height=0.75\textwidth,
        xticklabel style = {font =\fontsize{\figureFontSize pt}{10pt}\selectfont},
        yticklabel style = {font=\fontsize{\figureFontSize pt}{10pt}\selectfont},
        xlabel style = {font =\fontsize{\figureFontSize pt}{\figureFontSize pt}\selectfont},
        ylabel style = {font=\fontsize{\figureFontSize pt}{\figureFontSize pt}\selectfont},
        legend columns=2,
        legend style={at={(0,-0.2)},anchor=north west,
        nodes={font=\fontsize{\figureFontSize pt}{\figureFontSize pt}\selectfont}}
        ]
    
    \addplot[color = red, dashed, line width=0.75pt] table[x=stp, y=t_je, col sep=comma] {Tikz_figures/bracket_steps_time.csv};
    
    \addplot[color = red, densely dotted, mark = *, line width=0.75pt] table[x=stp, y=t_bje, col sep=comma] {Tikz_figures/bracket_steps_time.csv};

    \addlegendentry{Jacobi enriched}
    \addlegendentry{Block Jacobi enriched}
   
    \end{axis}

\end{tikzpicture}
	\end{subfigure}
	
	\caption{Notched bearing bracket. Number of iterations and wall time at each step of crack propagation for the alternatives considered using 5,000 subdomains.}
	\label{fig:bracket_steps_iterations}
\end{figure}

Similar to the previous example, the final test for this example involves a propagating crack. In \autoref{fig:bracket_cracks_deformed}, the deformed mesh is illustrated for the initial crack and after 15 crack propagation steps. In \autoref{fig:bracket_steps_iterations} the number of iterations and wall time required at each crack propagation step are given for the two alternatives considered using 5,000 subdomains. As shown in the previous paragraph, the proposed block Jacobi preconditioner is much more efficient for this example. Furthermore, in both cases the number of iterations, and, as a result, the computational time required, remains almost unaffected as the crack propagates.

\subsection{Cracked compressor blade}

\begin{figure}
	\centering
	\begin{subfigure}[t]{0.30\textwidth}
	    \centering
		\includegraphics[width=0.5\textwidth]{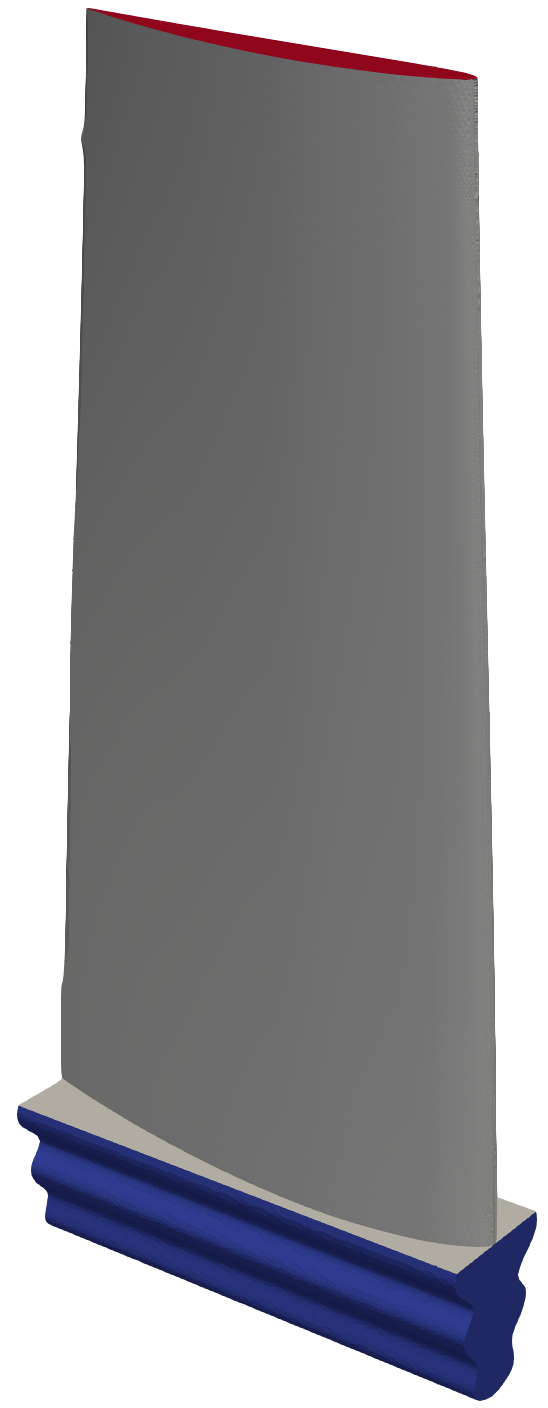}
	    \caption{Cracked compressor blade geometry and boundary conditions. Blue colour corresponds to fixed displacements and red to a uniform traction in the vertical direction.}
	    \label{fig:blade_bcs}
	\end{subfigure} \qquad
	\begin{subfigure}[t]{0.43\textwidth}
		\includegraphics[width=1\textwidth]{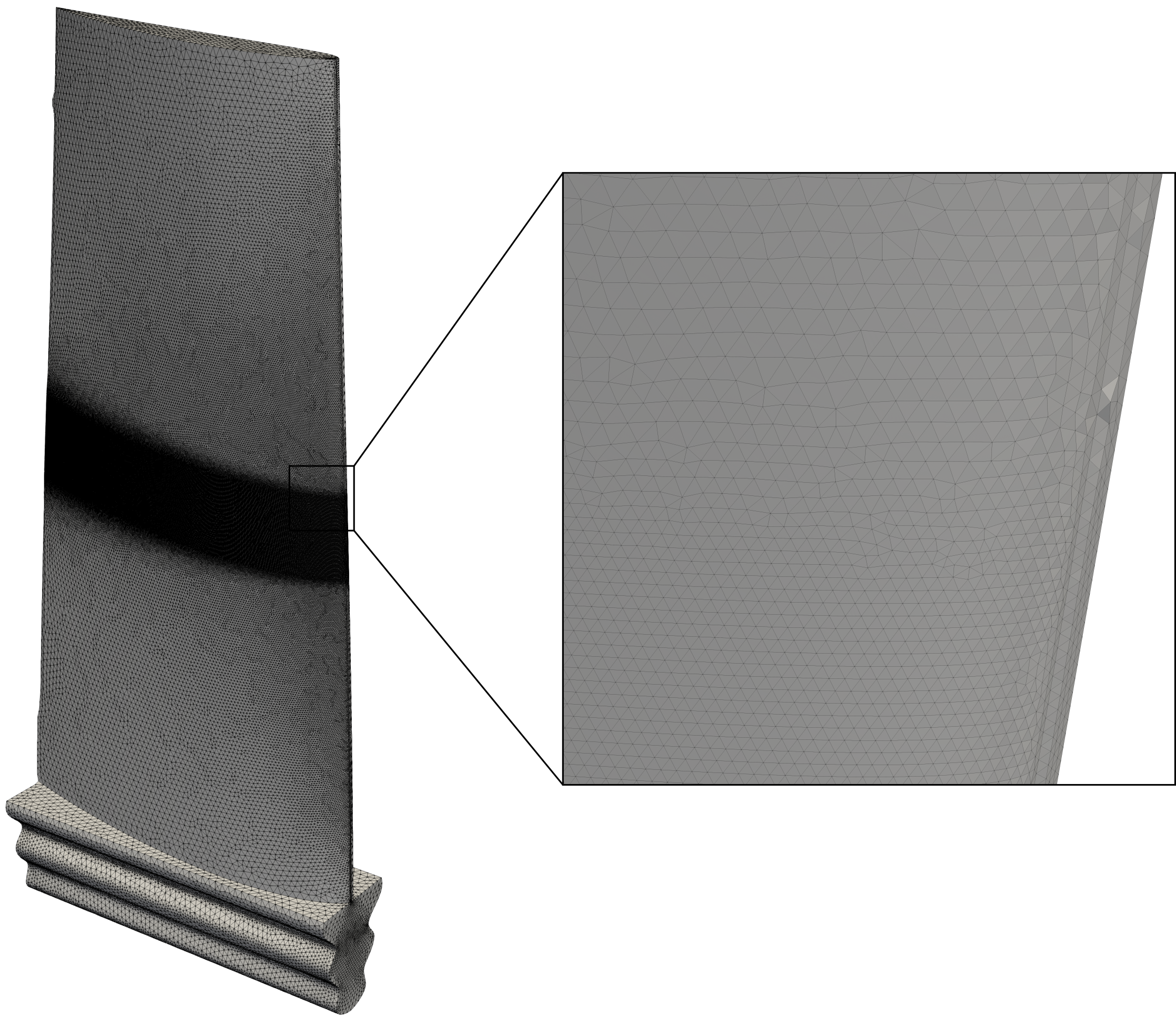}
	    \caption{Cracked compressor blade mesh.}
	    \label{fig:blade_mesh}
	\end{subfigure}
	\caption{Cracked compressor blade, mesh, geometry and boundary conditions.}
	\label{fig:blade}
\end{figure}

In the final example, the geometry of a high-pressure compressor blade, obtained from \href{https://grabcad.com/library/rotor-blade-5}{grabcad.com}, is considered, as illustrated in \autoref{fig:blade}. The material properties used are: $E=2.1 \times 10^5$ N/mm$^2$, $\nu=0.3$, while boundary conditions applied are illustrated in \autoref{fig:blade_bcs}, where blue areas correspond to fixed displacements and the red one to uniform tractions in the vertical direction. 

The initial crack considered is circular 1mm crack, introduced at the location illustrated in \autoref{fig:blade_initial_crack}. An unstructured mesh with size in the range $h=0.2-2$mm is used, as illustrated in \autoref{fig:blade_mesh}, whose statistics are given in \autoref{tab:blade_mesh_stats}. The smaller mesh size is used in the area around the crack to accurately represent its geometry and resulting stress, strain and displacement fields, as well as at other locations of the blade to represent smaller geometrical features. For the enriched part, the enrichment radius is set to $r_e=0.2$mm since it was observed that higher enrichment radii led to fluctuations of the crack surface along the boundaries of the domain.

\begin{table}
\centering
\caption{Cracked compressor blade. Mesh statistics for the mesh used. The number of enriched dofs refers to the initial configuration, before any crack propagation occurs.}
\begin{tabular}{ |l| c|}
\hline
              &  $h = 0.2-2$ mm   \\
\hline
elements      &  10,890,279  \\
nodes         &  1,825,964  \\
dofs          &  5,477,892  \\
enriched dofs &  876  \\
\hline
\end{tabular}
\label{tab:blade_mesh_stats}
\end{table}

\subsubsection{Effectiveness and robustness of the proposed preconditioning scheme for enriched approximations}

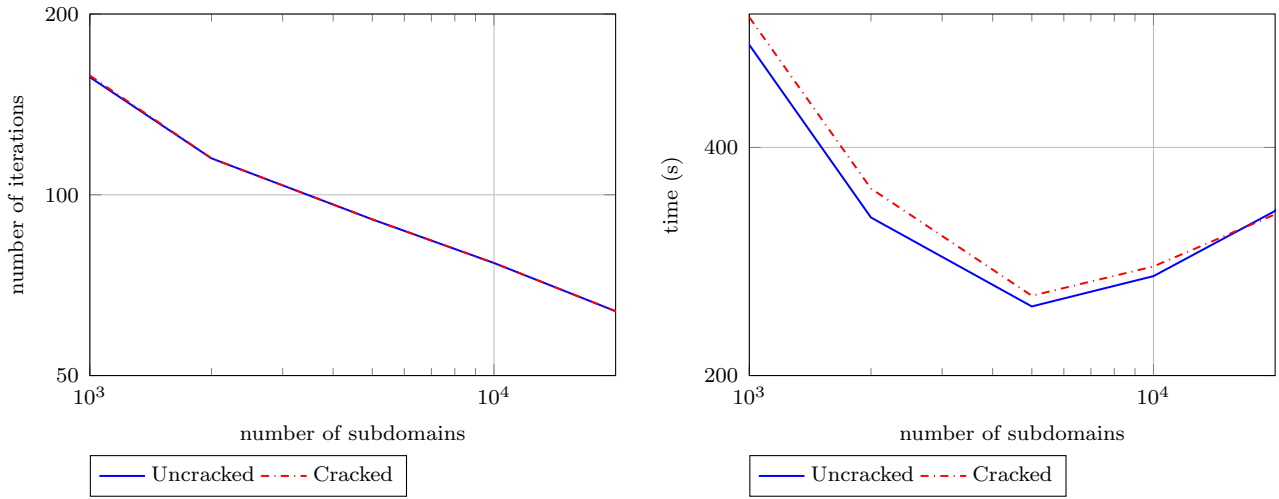
\begin{figure}
	\begin{subfigure}[b]{0.5\textwidth}
		\begin{tikzpicture}
    \begin{axis}[
        xmin = 1000,
        xmax = 20000,
        ymin = 50,
        ymax = 200,
        xtick = {1000,10000},
        xticklabels = {$10^3$,$10^4$},
        ytick = {50,100,200},
        yticklabels = {50,100,200},
        xmode = log,
        ymode = log,
        xlabel = {number of subdomains},
        ylabel = {number of iterations},
        grid = both,
        xminorgrids=false,
        yminorgrids=false,
        width=1\textwidth,
        height=0.75\textwidth,
        xticklabel style = {font =\fontsize{\figureFontSize pt}{10pt}\selectfont},
        yticklabel style = {font=\fontsize{\figureFontSize pt}{10pt}\selectfont},
        xlabel style = {font =\fontsize{\figureFontSize pt}{\figureFontSize pt}\selectfont},
        ylabel style = {font=\fontsize{\figureFontSize pt}{\figureFontSize pt}\selectfont},
        legend columns=2,
        legend style={at={(0,-0.22)},anchor=north west,
        nodes={font=\fontsize{\figureFontSize pt}{\figureFontSize pt}\selectfont}}
        ]
    
    \addplot[color = blue,line width=0.75pt] table[x=sd, y=it, col sep=comma] {Tikz_figures/blade_subdomain_it_t.csv};
    
    \addplot[color = red, dashdotted,line width=0.75pt] table[x=sd, y=it_c, col sep=comma] {Tikz_figures/blade_subdomain_it_t.csv};
    

   	\addlegendentry{Uncracked}
   	\addlegendentry{Cracked}
   
    \end{axis}
\end{tikzpicture}
	\end{subfigure}
	\begin{subfigure}[b]{0.5\textwidth}
		\begin{tikzpicture}
    \begin{axis}[
        xmin = 1000,
        xmax = 20000,
        ymin = 200,
        ymax = 600,
        xtick = {1000,10000},
        xticklabels = {$10^3$,$10^4$},
        ytick = {200,400},
        yticklabels = {200,400},
        xmode = log,
        ymode = log,
        xlabel = {number of subdomains},
        ylabel = {time (s)},
        grid = both,
        xminorgrids=false,
        yminorgrids=false,
        width=1\textwidth,
        height=0.75\textwidth,
        xticklabel style = {font =\fontsize{\figureFontSize pt}{10pt}\selectfont},
        yticklabel style = {font=\fontsize{\figureFontSize pt}{10pt}\selectfont},
        xlabel style = {font =\fontsize{\figureFontSize pt}{\figureFontSize pt}\selectfont},
        ylabel style = {font=\fontsize{\figureFontSize pt}{\figureFontSize pt}\selectfont},
        legend columns=2,
        legend style={at={(0,-0.22)},anchor=north west,
        nodes={font=\fontsize{\figureFontSize pt}{\figureFontSize pt}\selectfont}}
        ]
    
    \addplot[color = blue,line width=0.75pt] table[x=sd, y=t, col sep=comma] {Tikz_figures/blade_subdomain_it_t.csv};
    
    \addplot[color = red, dashdotted,line width=0.75pt] table[x=sd, y=t_c, col sep=comma] {Tikz_figures/blade_subdomain_it_t.csv};
    

   	\addlegendentry{Uncracked}
   	\addlegendentry{Cracked}
   
    \end{axis}
\end{tikzpicture}
	\end{subfigure}
	\caption{Cracked compressor blade. Performance of deflation in conjunction to a block Jacobi preconditioner for the cracked and uncracked blade.}
	\label{fig:blade_subdomain_performance}
\end{figure}

For this example, only results for the block Jacobi preconditioner are reported in \autoref{fig:blade_subdomain_performance}, in terms of the number of iterations and wall time required for the blade without and with the crack. The behaviour observed is slightly different from the previous examples, since the number of iterations seems to be decreasing throughout the full range of subdomain numbers tested, while wall time initially decreases and subsequently increases again. This can be attributed to a number of factors. Firstly, the time reported includes the time for partitioning the domain as well as the time for initialising the preconditioner, both of which are quite substantial for this example and can affect the overall timing. In fact, the time required for the solution itself is significantly smaller and keeps decreasing for the 10,000 subdomain case, as will be shown in the following paragraph. Additionally, for the numbers of subdomains required for this example, the size of the linear systems of Algorithms \ref{alg:A-DEF2_u0} and \ref{alg:A-DEF2} becomes quite substantial. Since these linear systems have to be solved at each iteration, they can slow down the preconditioner, especially as the number of subdomains increases.

In the present example, the numbers of iterations required for the uncracked and cracked case are identical, since, in a similar fashion to the previous example, the crack is very small compared to the domain considered and affects the solution only locally. The differences in wall time observed, can be attributed to the fact that, in our implementation, the preconditioner is initialised without considering the crack and then updated at each crack propagation step by factorising the blocks of the stiffness matrix corresponding to enriched subdomains.

\subsubsection{Performance of the proposed scheme for propagating cracks}

\begin{figure}
    \centering
    \begin{subfigure}[b]{0.175\textwidth}
		\includegraphics[width=1.0\textwidth]{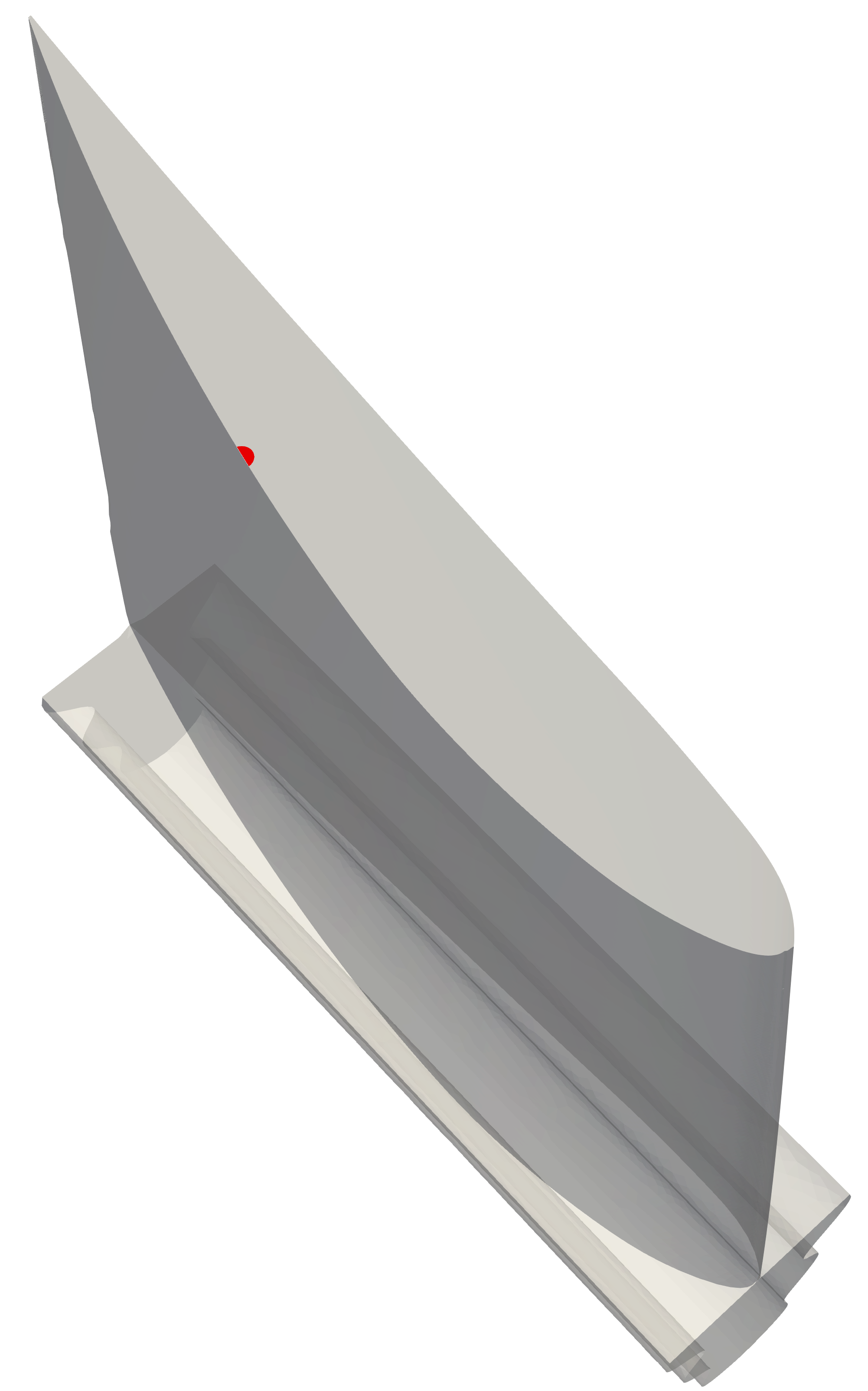}
		\caption{Initial crack highlighted in red.}\label{fig:blade_initial_crack}
	\end{subfigure} \qquad
	\begin{subfigure}[b]{0.19\textwidth}
		\includegraphics[width=1.0\textwidth]{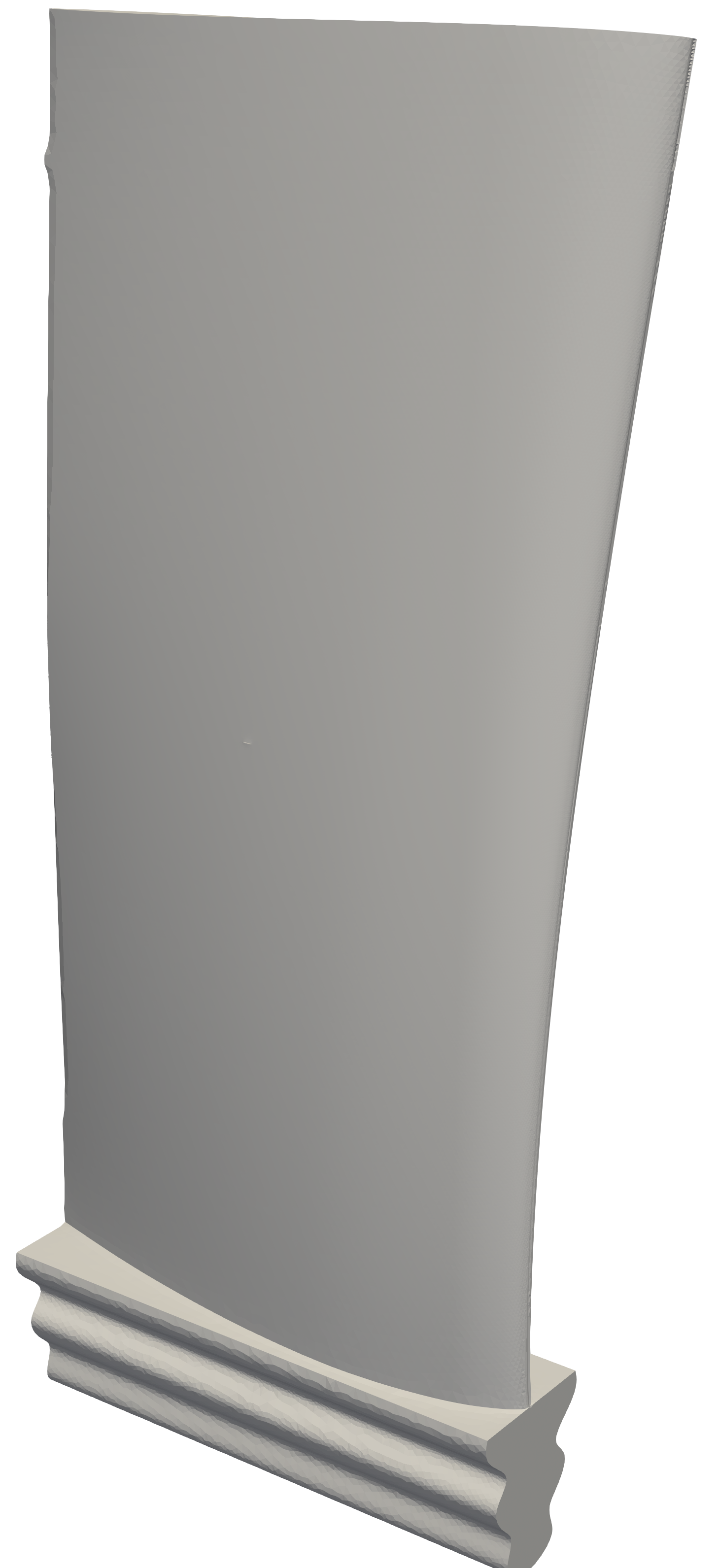}
		\caption{Initial crack deformed mesh.}\label{fig:blade_initial_crack_deformed}
	\end{subfigure} \qquad
	\begin{subfigure}[b]{0.175\textwidth}
		\includegraphics[width=1.0\textwidth]{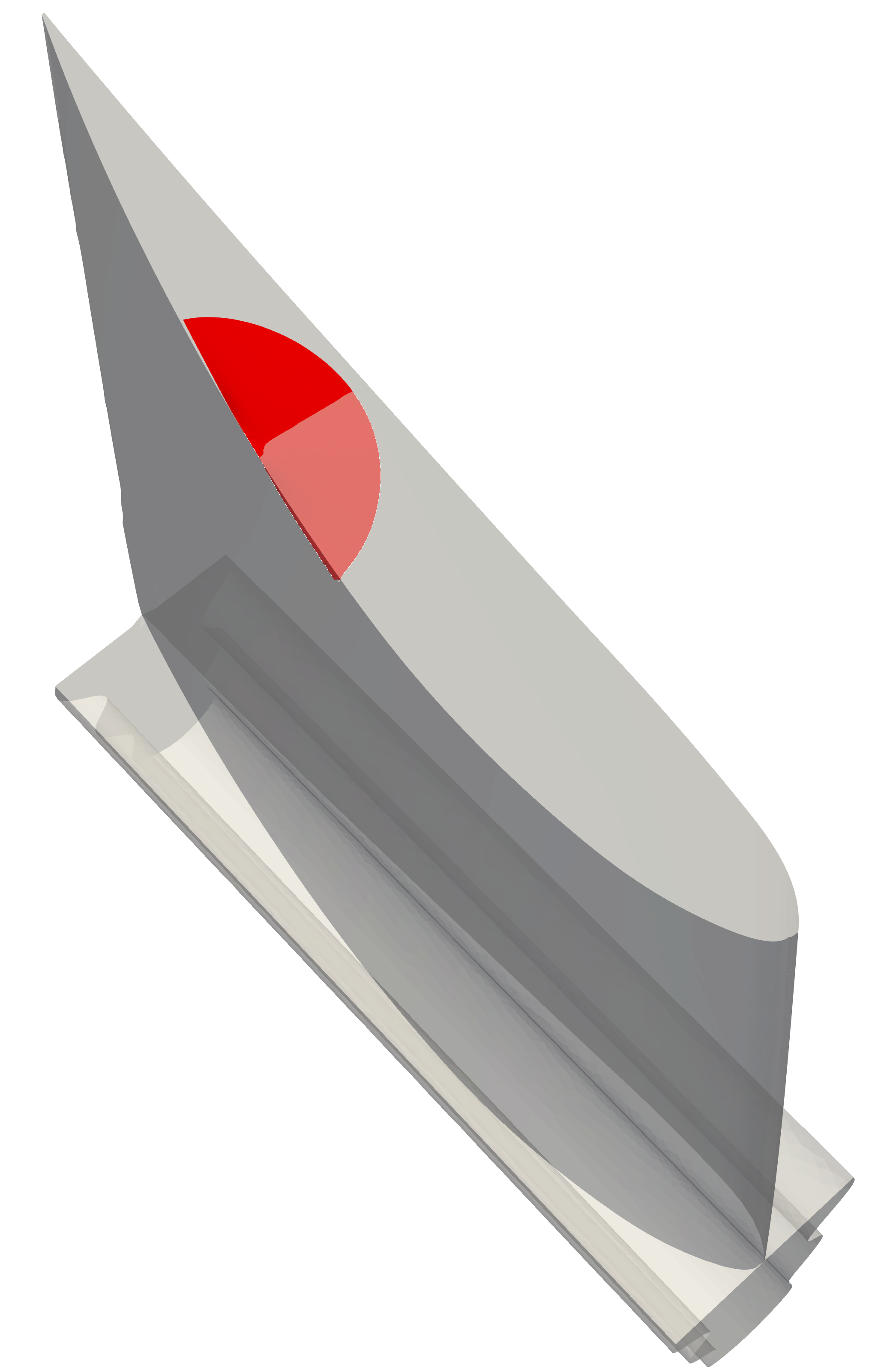}
		\caption{Final crack highlighted in red.}\label{fig:blade_final_crack}
	\end{subfigure} \qquad
	\begin{subfigure}[b]{0.20\textwidth}
		\includegraphics[width=1.0\textwidth]{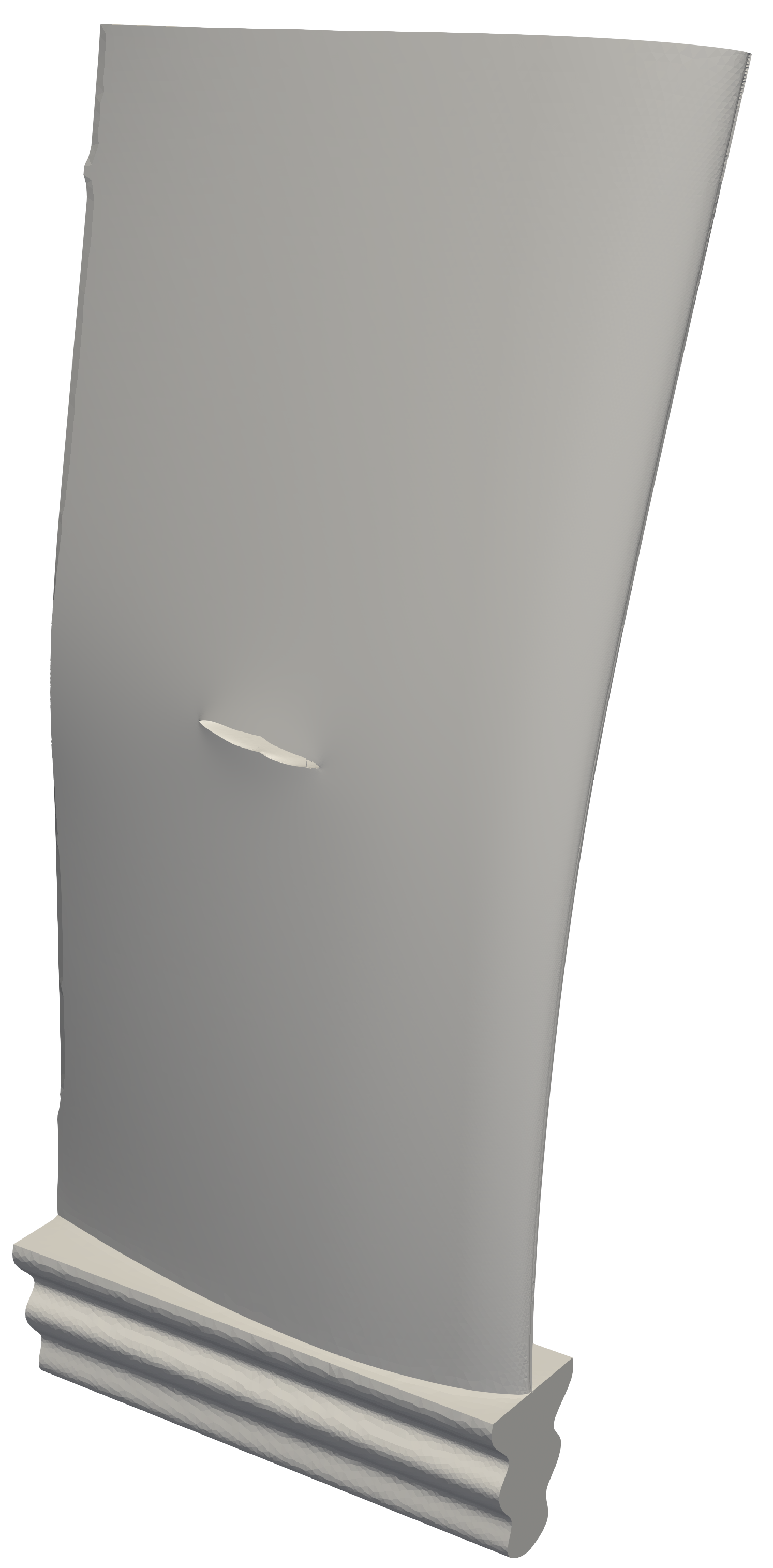}
		\caption{Final crack deformed mesh.}\label{fig:blade_final_crack_deformed}
	\end{subfigure}
	\caption{Cracked compressor blade. Section with highlighted crack surface and deformed mesh for the initial and propagated crack after 60 steps of crack propagation.}
	\label{fig:blade_cracks_deformed}
\end{figure}

As a final test the initial 1mm crack, illustrated in \autoref{fig:blade_initial_crack}, is propagated for 60 steps, until it almost penetrates the blade, as shown in \autoref{fig:blade_final_crack}. The deformed blade is also illustrated for the initial and final state in Figures \ref{fig:blade_initial_crack_deformed} and \ref{fig:blade_final_crack_deformed} respectively, where it can be noticed that the propagated crack substantially alters the response of the blade. For this test, 10,000 subdomains were used requiring between 76 and 77 iterations and 177s to 178s per solution throughout the simulation. Similar to the previous example, both the number of iterations and the wall time required were only minimally affected by crack propagation. 

\section{Conclusions and discussion} \label{conclusion}

A preconditioning strategy was proposed to accelerate the iterative solution of linear systems resulting from the discretization of fracture mechanics problems with extended/generalised finite elements. The approach involves deflation and a block Jacobi preconditioner combined in a multiplicative way. The deflation space typically used for linear elasticity problems is enriched to account for discontinuities and singularities introduced in the solution by enrichment, thus effectively removing low frequency components of the error. The block Jacobi preconditioner is used to remove high frequency components of the error, as well as linear dependencies associated with enrichment, by accounting for interactions between different dofs.

A series of numerical studies were conducted to test the performance of the proposed scheme, where the following trends were observed:

\begin{itemize}
    \item Enriching the deflation space consistently improves the performance of the preconditioner both in terms of iterations required and computational time.
    \item The proposed block Jacobi preconditioner effectively removes linear dependencies introduced by enrichment, leading to substantial improvements in iteration numbers and computational time when compared to a Jacobi preconditioner, which is a common choice for deflated CG methods.
    \item For the smaller systems tested ($\approx$500,000 dofs), the approach outperforms state of the art direct solvers. Especially for crack propagation, where the preconditioner can be set up once at the beginning of the simulation, this difference in performance can be quite substantial.
    \item For the larger systems tested, where cracks are small compared to the size of the components considered, the number of iterations and commputational time required at each crack propagation step remains almost unaffected as cracks propagate.
\end{itemize}

While all of the methods employed in the current work are very well suited for parallel computing, the current implementation only makes limited use of this potential. Therefore, as a direction of future work, a fully optimised implementation of the methods will be developed, utilising distributed memory parallelism, to improve performance. Moreover, since the current approach is essentially a two level preconditioner, future work will focus on extending to a multi-level method, in order to improve scalability and allow the solution of larger problems. Finally, the possibility of re-using solutions obtained in previous crack propagation steps to accelerate the solution of subsequent steps will be explored.

\bibliography{references.bib}

\end{document}